\documentclass[longauth]{aa}  

\usepackage{graphicx}
\usepackage{txfonts}
\usepackage{amssymb}
\usepackage{amsmath}
\usepackage{graphicx}
\usepackage[colorlinks=true, allcolors=blue]{hyperref}
\usepackage{booktabs}
\usepackage{amsfonts}     
\usepackage[T1]{fontenc} 
\usepackage[utf8]{inputenc} 
\usepackage{multicol}
\usepackage{lineno}
\usepackage{gensymb}
\usepackage{soul}
\usepackage{rotating}
\usepackage{comment}
\usepackage{bm}
\usepackage[dvipsnames]{xcolor}
\usepackage{float}

\begin{document}

   \title{Time-dependent modelling of short-term variability in the TeV-blazar VER~J0521+211 during the major flare in 2020}
   
   \author{
MAGIC Collaboration: S.~Abe\inst{1} \and
J.~Abhir\inst{2} \and
A.~Abhishek\inst{3} \and
V.~A.~Acciari\inst{4} \and
A.~Aguasca-Cabot\inst{5} \and
I.~Agudo\inst{6} \and
T.~Aniello\inst{7} \and
S.~Ansoldi\inst{8,41} \and
L.~A.~Antonelli\inst{7} \and
A.~Arbet Engels\inst{9}
% ...
}

    \institute { Japanese MAGIC Group: Institute for Cosmic Ray Research (ICRR), The University of Tokyo, Kashiwa, 277-8582 Chiba, Japan
\and ETH Z\"urich, CH-8093 Z\"urich, Switzerland
\and Universit\`a di Siena and INFN Pisa, I-53100 Siena, Italy
\and Institut de F\'isica d'Altes Energies (IFAE), The Barcelona Institute of Science and Technology (BIST), E-08193 Bellaterra (Barcelona), Spain
\and Universitat de Barcelona, ICCUB, IEEC-UB, E-08028 Barcelona, Spain
\and Instituto de Astrof\'isica de Andaluc\'ia-CSIC, Glorieta de la Astronom\'ia s/n, 18008, Granada, Spain
\and National Institute for Astrophysics (INAF), I-00136 Rome, Italy
\and Universit\`a di Udine and INFN Trieste, I-33100 Udine, Italy
\and Max-Planck-Institut f\"ur Physik, D-85748 Garching, Germany
\and Universit\`a di Padova and INFN, I-35131 Padova, Italy
\and Technische Universit\"at Dortmund, D-44221 Dortmund, Germany
\and Croatian MAGIC Group: University of Zagreb, Faculty of Electrical Engineering and Computing (FER), 10000 Zagreb, Croatia
\and Centro Brasileiro de Pesquisas F\'isicas (CBPF), 22290-180 URCA, Rio de Janeiro (RJ), Brazil
\and IPARCOS Institute and EMFTEL Department, Universidad Complutense de Madrid, E-28040 Madrid, Spain
\and Instituto de Astrof\'isica de Canarias and Dpto. de  Astrof\'isica, Universidad de La Laguna, E-38200, La Laguna, Tenerife, Spain
\and University of Lodz, Faculty of Physics and Applied Informatics, Department of Astrophysics, 90-236 Lodz, Poland
\and Centro de Investigaciones Energ\'eticas, Medioambientales y Tecnol\'ogicas, E-28040 Madrid, Spain
\and Departament de F\'isica, and CERES-IEEC, Universitat Aut\`onoma de Barcelona, E-08193 Bellaterra, Spain
\and Universit\`a di Pisa and INFN Pisa, I-56126 Pisa, Italy
\and INFN MAGIC Group: INFN Sezione di Bari and Dipartimento Interateneo di Fisica dell'Universit\`a e del Politecnico di Bari, I-70125 Bari, Italy
\and Department for Physics and Technology, University of Bergen, Norway
\and INFN MAGIC Group: INFN Sezione di Torino and Universit\`a degli Studi di Torino, I-10125 Torino, Italy
\and Croatian MAGIC Group: University of Rijeka, Faculty of Physics, 51000 Rijeka, Croatia
\and Universit\"at W\"urzburg, D-97074 W\"urzburg, Germany
\and Japanese MAGIC Group: Physics Program, Graduate School of Advanced Science and Engineering, Hiroshima University, 739-8526 Hiroshima, Japan
\and Deutsches Elektronen-Synchrotron (DESY), D-15738 Zeuthen, Germany
\and Armenian MAGIC Group: ICRANet-Armenia, 0019 Yerevan, Armenia
\and Croatian MAGIC Group: Josip Juraj Strossmayer University of Osijek, Department of Physics, 31000 Osijek, Croatia
\and Finnish MAGIC Group: Finnish Centre for Astronomy with ESO, Department of Physics and Astronomy, University of Turku, FI-20014 Turku, Finland
\and Japanese MAGIC Group: Department of Physics, Tokai University, Hiratsuka, 259-1292 Kanagawa, Japan
\and University of Geneva, Chemin d'Ecogia 16, CH-1290 Versoix, Switzerland
\and Saha Institute of Nuclear Physics, A CI of Homi Bhabha National Institute, Kolkata 700064, West Bengal, India
\and Inst. for Nucl. Research and Nucl. Energy, Bulgarian Academy of Sciences, BG-1784 Sofia, Bulgaria
\and Japanese MAGIC Group: Department of Physics, Yamagata University, Yamagata 990-8560, Japan
\and Finnish MAGIC Group: Space Physics and Astronomy Research Unit, University of Oulu, FI-90014 Oulu, Finland
\and Japanese MAGIC Group: Chiba University, ICEHAP, 263-8522 Chiba, Japan
\and Japanese MAGIC Group: Institute for Space-Earth Environmental Research and Kobayashi-Maskawa Institute for the Origin of Particles and the Universe, Nagoya University, 464-6801 Nagoya, Japan
\and Japanese MAGIC Group: Department of Physics, Kyoto University, 606-8502 Kyoto, Japan
\and INFN MAGIC Group: INFN Roma Tor Vergata, I-00133 Roma, Italy
\and Japanese MAGIC Group: Department of Physics, Konan University, Kobe, Hyogo 658-8501, Japan
\and also at International Center for Relativistic Astrophysics (ICRA), Rome, Italy
% ...
}

   \date{Received xxx; accepted yyy}

 
  \abstract{The BL Lacertae object VER~J0521+211 underwent a notable flaring episode in February 2020. A short-term monitoring campaign, led by the MAGIC (Major Atmospheric Gamma Imaging Cherenkov) collaboration, covering a wide energy range from radio to very-high-energy (VHE, 100 GeV $<$ $E$ $<$ 100 TeV) gamma rays was organised to study its evolution. These observations resulted in a consistent detection of the source over six consecutive nights in the VHE gamma-ray domain. Combining these nightly observations with an extensive set of multiwavelength data made modelling of the blazar's spectral energy distribution (SED) possible during the flare. This modelling was performed with a focus on two plausible emission mechanisms: i) a leptonic two-zone synchrotron-self-Compton scenario, and ii) a lepto-hadronic one-zone scenario. Both models effectively replicated the observed SED from radio to the VHE gamma-ray band. Furthermore, by introducing a set of evolving parameters, both models were successful in reproducing the evolution of the fluxes measured in different bands throughout the observing campaign. Notably, the lepto-hadronic model predicts enhanced photon and neutrino fluxes at ultra-high energies ($E$ $>$ $100\,\mathrm{TeV}$). While the photon component, generated via decay of neutral pions, is not directly observable as it is subject to intense pair production (and therefore extinction) through interactions with the cosmic microwave background photons, neutrino detectors (e.g. IceCube) can probe the predicted neutrino component. Finally, the analysis of the gamma-ray spectra, as observed by MAGIC and the {\it Fermi}-LAT telescopes, yielded a conservative 95\% confidence upper limit of $z \leq 0.244$ for the redshift of this blazar.}

   
   \titlerunning{Time-dependent modelling of VER~J0521+211 during the major flare in 2020}
   \authorrunning{MAGIC Coll. et al.}

   \maketitle
%

\section{Introduction}
\label{sec:intro}

Blazars constitute a subclass of active galactic nuclei (AGN) whose most distinctive feature is a prominent jet, populated with highly relativistic particles, closely aligned with the observer's line-of-sight. This unique orientation causes the non-thermal radiation produced in the jet, which spans from radio frequencies to the very-high-energy (VHE, 100 GeV $<$ $E$ $<$ 100 TeV) gamma-ray band, to be heavily Doppler-boosted, often outshining the entire host galaxy. The broad-band spectral energy distribution (SED) of blazars exhibits a characteristic two-bump structure \citep{10.1093/mnras/stx806} with a first component (often known as the synchrotron bump after the emission process that is thought to be its origin) peaking between the infrared and X-ray frequencies, whilst the second component peak (with several competing processes possibly contributing to it) is located beyond the X-ray band. Further sub-classification of blazars has been suggested  \citep{urry1995unified} based on their synchrotron peak frequency, as well as the width and intensity of the emission lines that may be present in the blazar's optical spectrum.

VER~J0521+211 is a BL Lac object, often regarded as a borderline intermediate-frequency synchrotron-peaked BL Lac (IBL) to high-frequency synchrotron-peaked BL Lac (HBL), depending on whether the source is found in quiescent or elevated emission states, respectively.
Situated at right ascension 05h 21m 45.9s (J2000) and declination +21$^\circ$ 12' 51'' (J2000), VER~J0521+211 is associated with the high-energy (HE,  30 MeV $<$ $E$ $<$ 100 GeV) gamma-ray source 4FGL J0521.7+2112 in the fourth {\it Fermi}-LAT source catalogue \citep[4FGL-DR3,][]{abdollahi2020fermi}. In the latest version of the catalogue \citep[4FGL-DR4,][]{2022ApJS..260...53A}, which includes 14 years of data, the average flux of HE gamma rays for this source is $(1.162 \pm 0.023) \times 10^{-8}\,\mathrm{cm}^{-2}\mathrm{s}^{-1}$. The significance of the spectral curvature is above $5\,\sigma$, with a preferred log parabola spectral shape. Its amplitude is $(4.685\pm 0.077)\times 10^{-12}\,\mathrm{cm^{-2}s^{-1}MeV^{-1}}$, the spectral index $1.869\pm0.017$ and the curvature parameter $\beta=0.0437\pm0.0066$, all measured at a reference energy of $1565.24\,\mathrm{MeV}$.

VER~J0521+211 was initially detected at the VHE gamma-ray band by the VERITAS (Very Energetic Radiation Imaging Telescope Array System) collaboration in 2009. Subsequently, it was identified as spatially associated with the radio and X-ray source RGB J0521.8+2112 \citep{archambault2013discovery}. An enhanced emission period was observed in VHE gamma rays by both MAGIC (Major Atmospheric Gamma Imaging Cherenkov) and VERITAS \citep{veritas2022multiwavelength} in 2013 (the 2013 campaign, hereafter), following an extended period of enhanced HE gamma-ray emission identified by \textit{Fermi}-LAT. The observations during this period have been used to test both one- and two-zone leptonic emission component models \citep{2Comp}.

Following an unprecedented gamma-ray activity \citep{2020ATel13522....1Q}, the MAGIC telescopes conducted VHE gamma-ray observations of VER~J0521+211 between 26 February 2020 and 2 March 2020 (MJD 58905-58910, the 2020 campaign, hereafter). These observations were accompanied by an extensive multiwavelength (MWL) effort, which enabled us to characterise night by night the variable SED of the source from radio to VHE gamma rays.

The optical spectrum of VER~J0521+211 is dominated by continuum emission, typical for BL Lac-type sources, leading to substantial uncertainty concerning the redshift of this blazar. The various attempts to identify emission lines in the optical spectrum \citep{Shaw2013_Redshifts, archambault2013discovery, paiano2017redshift} remain inconclusive. Lower limits built over assumptions of the host galaxy luminosity \citep{paiano2017redshift} and upper bounds based on extrapolations of the HE gamma-ray spectrum to VHE gamma rays \citep{veritas2022multiwavelength} as well as modelling of the observed VHE gamma-ray spectrum \citep{sahu2023_redshift} set the current estimations for the source distance at $0.18 \lesssim z \lesssim 0.31$.

In this paper, we present the extensive MWL observations of VER~J0521+211 from the 2020 campaign, comparing its broadband SED with archival flux measurements and employing both to constrain potential leptonic and lepto-hadronic emission models. In Section~\ref{sec:obs&an}, we describe the utilised instruments, the data, and the corresponding analysis methodologies. In Sect.~\ref{sec:res}, we present the MWL characterisation of the source, the analysis of its spectrum, and the estimation of the source distance. In Sect.~\ref{sec:sed&modelling}, we report the results of the time-dependent modelling of the broadband emission of VER~J0521+211 during distinct phases of this flaring episode. In Sect.~\ref{sec:optpol}, we detail the results of the modelling of its long-term optical polarisation variability. Finally, in Sect.~\ref{sec:sum&con}, we summarise the most important findings of this study.

\section{Observations and data analysis}
\label{sec:obs&an}

\begin{figure*}
\centering
\includegraphics[width=1.0\textwidth]{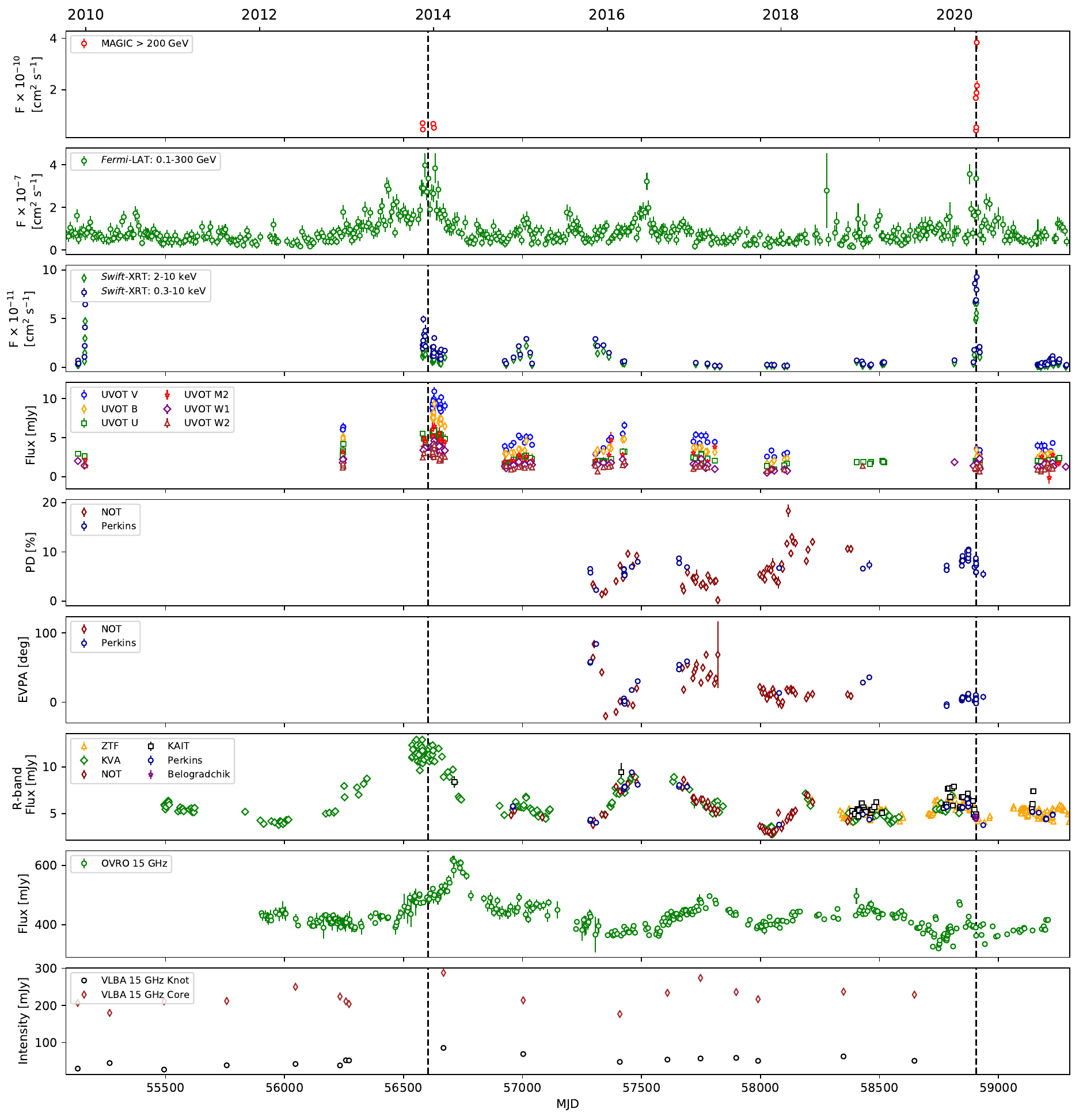}
\caption{Long-term MWL light curve of VER~0521+211 displaying the behaviour of the source from VHE gamma rays (top) to radio core intensity at 15 GHz (bottom). The dashed vertical lines indicate the two flaring episodes, 2013 flare as archival data from \citealt{2Comp} and the 2020 flare, in the VHE gamma-ray band.}
\label{Fig:longterm}
\end{figure*}

\begin{figure*}
\centering
\includegraphics[width=.86\textwidth]{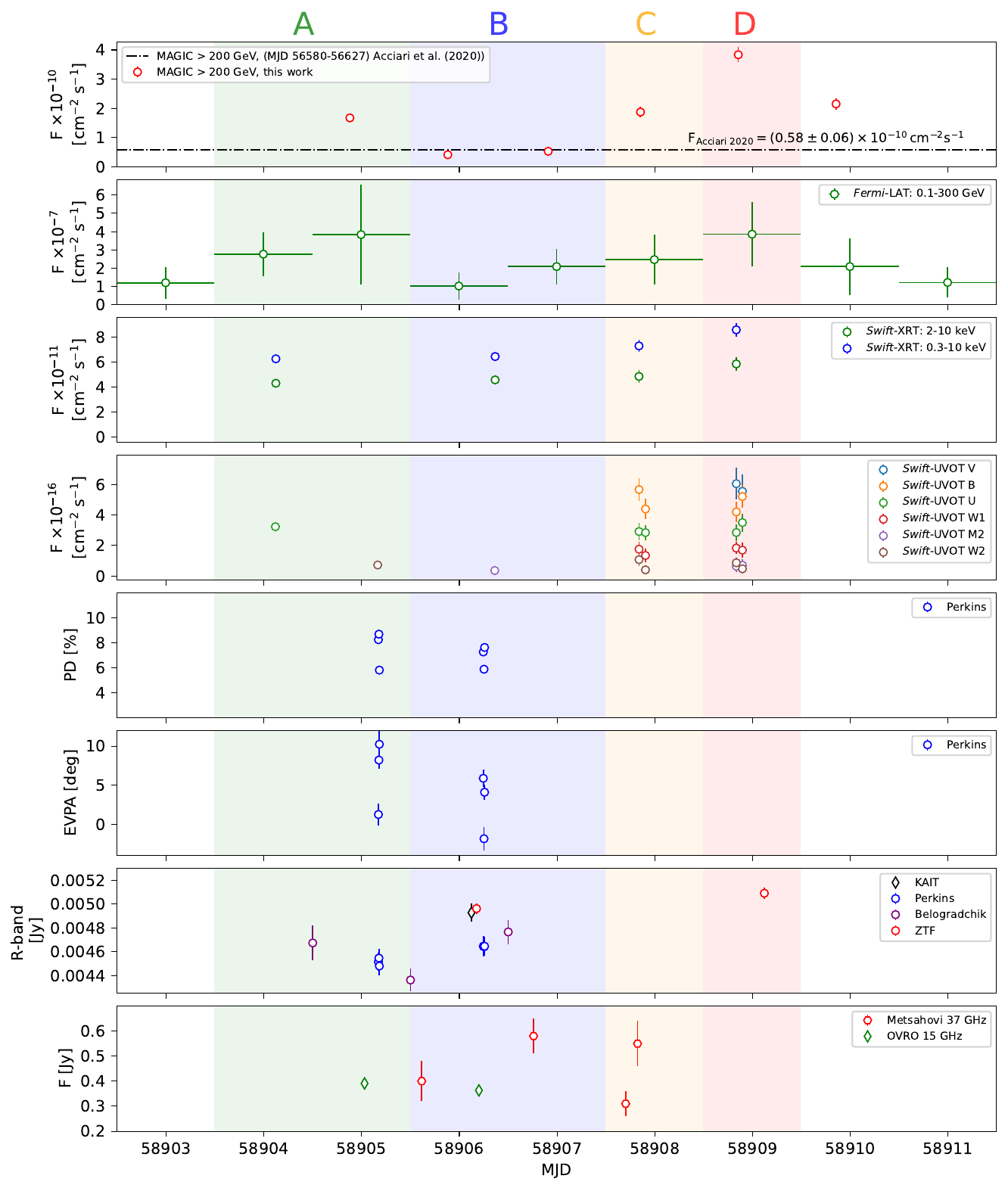}
\caption{Multiwavelength light curve showing the evolution of the flux of VER~J0521+211 in different bands and its optical polarisation during the 2020 flare. From top to bottom:
VHE gamma rays (with a reference value from \citealt{2Comp}), HE gamma-ray flux, X-ray flux, UV and optical in various bands, R-band flux from various optical telescopes as well as the evolution of the polarisation degree and electric vector polarisation angle, and radio flux. Shaded background colours define the states A-D used to build the SED of the source, as described in table \ref{tab:VHEspectra}.} 
\label{Fig:MWL_LC}
\end{figure*}

In the following subsections, we provide a brief introduction to each instrument and its corresponding data analysis procedure. Figure \ref{Fig:longterm} illustrates the long-term MWL data of VER~J0521+211, and Fig. \ref{Fig:MWL_LC} outlines the observations of the 2020 campaign during the flaring activity of the source. 
The MWL observations extend between 6 September 2009 (MJD 55080) and 27 March 2021 (MJD 59300) consisting of data from radio to VHE gamma rays.

\subsection{Very-high-energy gamma rays (MAGIC)}
\label{sec:magic}

MAGIC is a stereoscopic system comprising two imaging atmospheric Cherenkov telescopes, each consisting of a 17-m diameter mirror, situated at an elevation of 2200 meters above sea level on the Canary Island of La Palma (28.7$^{\circ}$ N, 17.9$^{\circ}$ W), Spain \citep{2016APh....72...76A}. The observations consisting of six consecutive nights were initiated on 26 February 2020  (MJD 58905) in response to an elevated state in the VHE gamma-ray emission of the source reported by the VERITAS collaboration \citep{2020ATel13522....1Q} and lasted until 2 March 2020 (MJD 58910).

The observations were done under diverse zenith angles, from 13 to 55 degrees, and night sky brightness (NSB) conditions, including periods of dark sky (3.9 hours) and moonlit conditions (2.1 hours). The data were analysed using the MAGIC standard analysis software \citep[MARS,][]{2013ICRC...33.2937Z}. The presence of moonlight affects the performance of the MAGIC telescopes, thereby impacting the subsequent data analysis and requiring tailored simulations. The description of the analysis of moonlit data in MAGIC is covered in \cite{arXiv:1704.00906}. Moreover, atmospheric transmission corrections were implemented to account for the imperfect reconstruction of air shower images from passing high-altitude clouds. These corrections were based on the continuous data recorded with the light detection and ranging instrument \citep[LIDAR,][]{lidarfruck,lidarschmuckermaier} which is one of the sub-systems of the MAGIC telescopes.

The archival data from \cite{2Comp} in the VHE gamma-ray is presented for completion in Fig. \ref{Fig:longterm}, and the average flux in the VHE gamma-ray band for this period, $F_{E \ > \ 200\,\mathrm{GeV}} = (0.58 \pm  0.06) \times 10^{-11}\,\mathrm{cm^{-2}s^{-1}}$, is presented as a reference flux line in Fig~\ref{Fig:MWL_LC}.

\subsection{High-energy gamma rays ({\it Fermi}-LAT)}
\label{sec:fermilat}

The large area telescope \citep[LAT,][]{atwood2009large} onboard the {\it Fermi} satellite is a pair-conversion telescope equipped with a precision converter-tracker and a calorimeter, enabling the detection of HE gamma rays. In this work, we analysed public data of {\it Fermi}-LAT 
\citep{2023ApJS..265...31A}
during the flare (Fig. \ref{Fig:MWL_LC}). For completeness, a long-term multiwavelength light curve (see Fig. \ref{Fig:longterm}) shows the data across the whole period between June 2009 and March 2021 (MJD 55080-59300, see Sect. \ref{sec:res&mwlvar}). 

For the data reduction, we used the {\tt fermipy} package \citep{https://doi.org/10.48550/arxiv.1707.09551}. We selected photons within the Pass 8 SOURCE class data \citep{atwood2013pass}, confined within a circular region of interest (ROI) centred on the target source's coordinates (as reported in Sect. \ref{sec:intro}), with a radius of 15$^{\circ}$. Throughout the data processing, we employed the instrument response functions  ``P8R3\_SOURCE\_V2'', and performed a binned likelihood analysis with bin size of 0.08$^{\circ}$ in both sky direction, and 8 bins per decade in energy from $0.1$ to $300\,\mathrm{GeV}$. We corrected all sources in the ROI for energy dispersion. Additionally, we applied standard quality cuts (``DATA\_QUAL 0 \&\& LAT\_CONFIG==1''), alongside a zenith distance cut of Zd < 90$^{\circ}$ to mitigate Earth limb contamination. The skymodel for the analysis consisted of the standard galactic \citep{acero2016development} and isotropic diffuse emission models (``gll\_iem\_v06.fits'' and ``iso\_P8R3\_SOURCE\_V2\_v1.txt'' respectively), in addition to all sources listed in the 4FGL-DR3 catalogue \citep[4FGL-DR3,][]{abdollahi2020fermi} within a 15$^{\circ}$ radius from the ROI centre. During the fitting procedure, we allowed the parameters of both diffuse components to vary, along with the spectral parameters of sources within a 5$^{\circ}$ radius surrounding the source of interest. We held the remaining parameters for sources within the ROI fixed based on the values published in the 4FGL-DR3 catalogue.

For the short-term evolution of the flux, we focused on a time interval coinciding with MAGIC observations between 23 February and 03 March 2020 (MJD 58902-58911), and performed a daily-binned analysis, with each time bin centred at midnight at the MAGIC site.

\subsection{X-rays ({\it Swift}-XRT)}
\label{sec:xray}

VER~J0521+211 was observed  at the time of the flare between 25 February and 1 March 2020 (MJD 58904-58909) by the X-ray Telescope \citep[XRT,][]{2004SPIE.5165..201B} onboard the \textit{Swift} satellite. The \textit{Swift}-XRT instrument log, publicly accessible through SwiftXRLOG\footnote{\url{https://heasarc.gsfc.nasa.gov/W3Browse/swift/swiftxrlog.html}}, provided a multi-epoch events list for these observations. The exposure time of each epoch is between $\sim 500 - 1400$ s resulting in a total exposure time of approximately 1.1 h. Following the methodology outlined in \citet{2017A&A...608A..68F}, we processed the data assuming a constant equivalent Galactic hydrogen column density of $N_H=4.38 \times 10^{21}\,\mathrm{cm}^{-2}$ \citep[as reported in][]{2013MNRAS.431..394W}, and produced the daily-binned light curve in two energy bands: $0.3-2.0\,$keV and $2-10\,$keV (Fig. \ref{Fig:MWL_LC}). Long-term view of the X-ray data retrieved from \citet{2Comp} is included in Fig. \ref{Fig:longterm} between June 2009 and March 2021 (MJD 55080-59300) for completeness.

\subsection{Optical and UV}
\label{sec:optical&uv}

\subsubsection{Optical-UV ({\it Swift}-UVOT)}
\label{sec:uvot}

During the 2020 campaign, the Ultraviolet/Optical Telescope (UVOT) onboard the \textit{Swift} satellite \citep{poole2008photometric} conducted five observations of VER~J0521+211 between 25 February and 1 March 2020 (MJD 58904-58909), four of which were simultaneous to \textit{Swift}-XRT observations. A variety of optical (V, B, U) and UV (W1, W2, W2) filter combinations \citep{poole2008photometric,breeveld2010further} were used. Moreover, individual filter observations were conducted on three occasions. Additionally, we collected archival data from the optical and UV bands between October 2009 and March 2020 (MJD 55131-59283) to the long-term dataset (see Fig. \ref{Fig:longterm}).

We used the \texttt{uvotmaghist} task from the HEAsoft package (v6.28) with the 10 November 2020 release of the \textit{Swift}/UVOT calibration database\footnote{\url{https://heasarc.gsfc.nasa.gov/docs/heasarc/caldb/swift/docs/swift_caldbhistory.html}} in the analysis of these data. We extracted the counts associated with VER~J0521+211 using a circular region with a radius of $5''$ centred at the source location, and performed the background estimation using a nearby source-free circular region with a radius of $20''$. We corrected the UVOT fluxes for Galactic extinction following \citet{2009ApJ...690..163R} assuming $\mathrm{E(B-V)=0.605}$ \citep{schlafly2011measuring}.

\subsubsection{Optical photometry (Tuorla, Perkins, KAIT, NOT, Belogradchik)}
\label{sec:optical}

The optical R-band light curve during the flare (Fig. \ref{Fig:MWL_LC}) and the long-term light curve extending from September 2010 to March 2020 (MJD 55450-59300, Fig. \ref{Fig:longterm}) incorporate data from multiple facilities. These include the Tuorla blazar monitoring program \footnote{\url{https://users.utu.fi/kani/1m/index.html}} \citep{takalo2008}, the Boston University Blazar Group\footnote{\url{http://www.bu.edu/blazars/BEAM-ME.html}} \citep{jorstad2016}, the Nordic Optical Telescope (NOT), and the Belogradchik Observatory. Both Figs. \ref{Fig:MWL_LC} and \ref{Fig:longterm} also show data from the KAIT \textit{Fermi} AGN Light-curve Reservoir\footnote{\url{http://herculesii.astro.berkeley.edu/kait/agn/}} acquired with the clear filter and Zwicky Transient Facility (ZTF)\footnote{\url{https://irsa.ipac.caltech.edu/Missions/ztf.html}} acquired with r and g filters. We matched the KAIT and the ZTF data to R-band data from the Tuorla blazar monitoring program using simultaneous observations and calculated a multiplication factor to shift them to the same level. Most of the data from the Tuorla monitoring program are collected using the Kungliga Vetenskapsakademien (KVA) telescope at Observatorio del Roque de los Muchachos (ORM) on La Palma, Spain. The Boston University Blazar Group obtained their data using the 1.83-m Perkins telescope in Arizona (USA). The NOT, a 2.56-m telescope, is located at ORM. The KAIT \textit{Fermi} AGN Light-curve Reservoir data are collected using a 76-cm robotic Katzman Automatic Imaging Telescope at the Lick Observatory in California, USA. The ZTF camera is mounted on a 117-cm robotic telescope at the Palomar Observatory in California, USA.

The data obtained from the Tuorla monitoring program and the NOT telescope is analysed following \citet{nilsson2018}. Similarly, we analysed the flat-, bias-, and dark-reduced images from the Belogradchik Observatory using the same analysis procedure as the Tuorla and the NOT data. The analysis procedures for the data from the Boston University Blazar Group and the KAIT Reservoir are described in \citet{jorstad2016} and \citet{li2003}, respectively. The ZTF data analysis procedure is explained in \citet{masci2019}.

We collected additional multiband optical data from the time of the flare, and used these data in the SED modelling (see Sect. \ref{sec:sed&modelling}). These include contributions from the Belogradchik Observatory (V- and I-bands) and from the Perkins Observatory (B-, V-, and I-bands). We applied galactic extinction corrections to all data, and additionally, corrected R- and I-band data for the host galaxy emission. To estimate the host galaxy contribution to the measured source flux, we assumed an elliptical galaxy with an absolute magnitude of $M = -22.8$ and an effective radius of $8\,\mathrm{kpc}$ at a redshift of $z = 0.18$ (see Sect. \ref{sec:res&zest}), and estimated its flux within an aperture of $5''$.

\subsubsection{Optical polarisation (Perkins, NOT)}
\label{sec:pol}

For the long-term optical polarisation analysis (see Sect. \ref{sec:optpol}), we collected optical polarisation light curves of VER J0521+211 in the  optical R-band, extending over the period from September 2015 to May 2020 (MJD 57280-58980). Our sample includes the optical R-band polarisation data from \citep{2Comp}, observations exclusively made using the NOT telescope, and supplementary data gathered from the Boston University Blazar Group (see Fig. \ref{Fig:longterm}). The instrumental setup of the NOT data is described in \citet{2Comp}, and the data analysis procedure in \citet{hovatta2016} and in \citet{magiccoll2018}. The description of the instrumental setup and the data analysis of the Boston University data can be found in \citet{jorstad2016}.

\subsection{Radio (OVRO, Metsähovi, MOJAVE)}
\label{sec:radio}

VER J0521+211 has been monitored in the radio band by the Owens Valley Radio Telescope (OVRO) at 15 GHz, and in this study, we included data covering both the flare (Fig. \ref{Fig:MWL_LC}) and the long-term light curve  between December 2011 and December 2020 (MJD 55900-59210). We collected additional data around the time of the flare from the 13.7-meter-diameter Metsähovi radio telescope at 37 GHz. For the assessment of the the long-term behaviour, we used the data from \citet{lister2021} observed within the MOJAVE (Monitoring Of Jets in Active galactic nuclei with VLBA Experiments) Project\footnote{\url{https://www.cv.nrao.edu/MOJAVE/}} at 15 GHz (see Fig. \ref{Fig:longterm}). The detailed analysis methodologies for OVRO and Metsähovi telescopes are described in \citet{richards2011blazars} and \citet{terasranta1998fifteen}, respectively. The analysis of MOJAVE data is described in \citet{lister2018}.

\section{Results}
\label{sec:res}

Blazars frequently display variability on various timescales, sometimes originating from distinct underlying physical processes occurring within the source. Using the extensive multiwavelength data described in Sect. \ref{sec:obs&an}, we constructed a comprehensive picture of VER~J0521+211 that convincingly describes the different observed emission states of the source. We put special focus on investigating the correlation and the delay between different bands, the changes in the optical polarisation over time, and the possibility of having a significant hadronic contribution to the total broadband emission of the source in the SED modelling.

\subsection{Multiwavelength variability}
\label{sec:res&mwlvar}

In addition to focusing on the exceptional flaring events such as the one described in this paper, it is also important to try to put these singular events into context by looking at the long-term behaviour of these sources across various spectral bands. This is particularly important since for most blazars following the temporal evolution of the flux changes in the VHE gamma-ray regime can only be done during such periods of enhanced emission. Therefore a multiwavelength view can help us shed light on the behaviour of these sources during and outside the VHE gamma-ray outbursts. This is also helpful in judging whether the VHE flares exhibit common patterns that repeat from one flare to another or if they are different in nature.

\begin{figure}
\centering
\includegraphics[width=\linewidth]{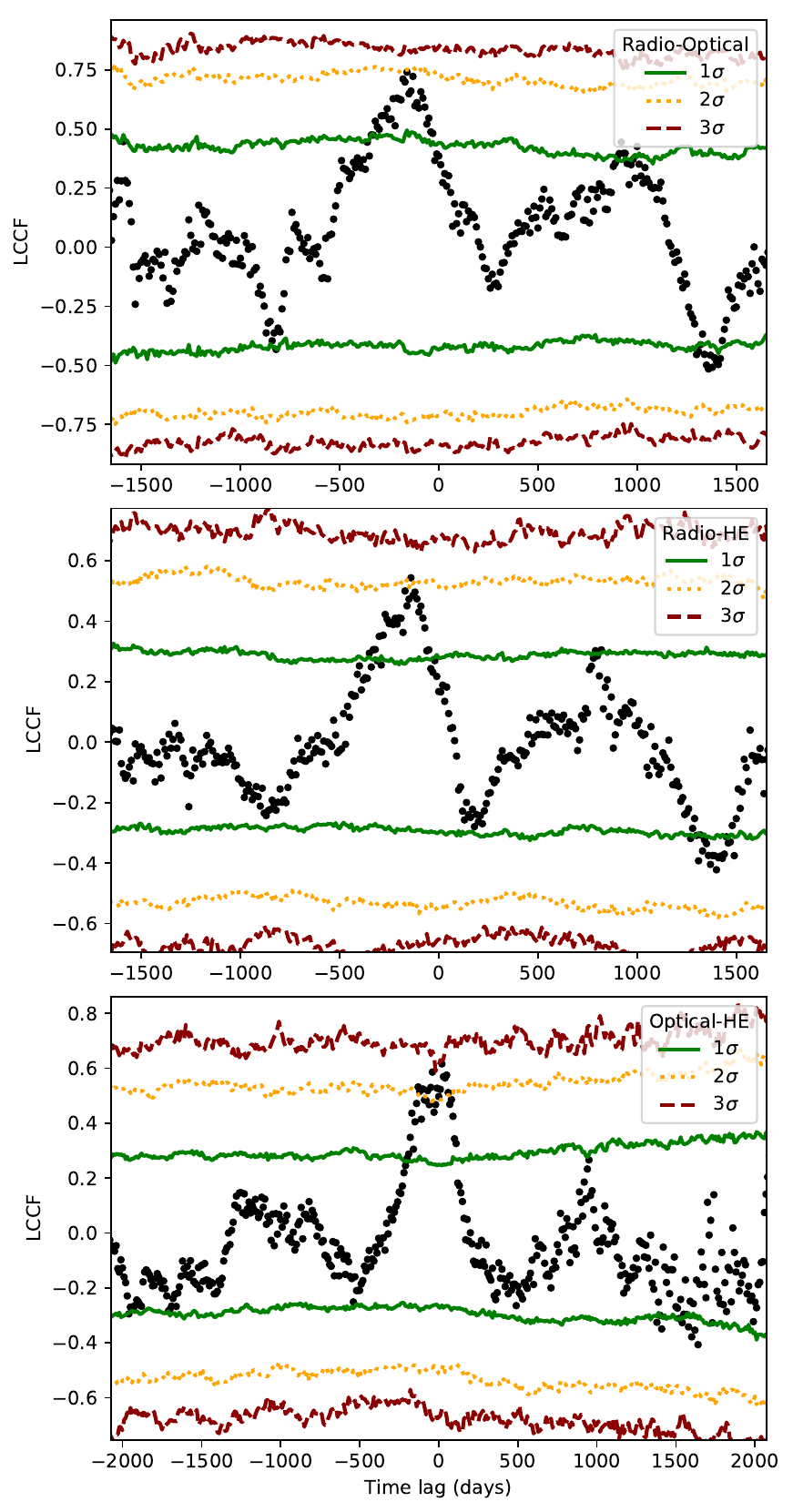}
\caption{DCF/LCCF correlation test results of the long-term radio and optical data, radio and HE gamma-ray data, and optical and HE gamma-ray data. The time lags found for each data set pair were $-160\pm11$, $-140\pm13$ and $20\pm12$ days. A negative time lag implies that the higher energy band is leading the lower energy band and a positive one the opposite. The solid green, dotted orange, and dashed red lines show the $1\sigma$, $2\sigma$, and $3\sigma$ significance levels respectively.}
\label{Fig:crosscorr}
\end{figure}

Using a discrete correlation function \citep[DCF,][]{1988ApJ...333..646E} with local normalisation \citep[LCCF,][]{1999PASP..111.1347W}, \citet{lindfors2016optical} carried out a cross-correlation analysis between the radio and optical bands covering their two-year data set. Due to the relatively short temporal integration, covering only a single rising edge, the cross-correlation analysis showed a plateau, only slightly favouring a time delay of about $110$ days, with the radio trailing behind the optical emission. Similar behaviour was found for the extended dataset in \cite{2Comp}. We repeated the same analysis using the long-term data detailed above (Fig. \ref{Fig:longterm}), covering now the rising and falling parts of several flare-like structures in the multiwavelength light curve. Our study revealed a clearer peak reaching $2\,\sigma$ confidence level in the DCF (see Fig. \ref{Fig:crosscorr}), corresponding to a time lag of $-160$ days, the optical leading the radio, with a standard deviation of 11 days. We estimated the significance levels of the correlation and the standard deviation of the time lag using a thousand simulated red-noise light curves with a power-law slope of -1.5 (optical) and -1.7 (radio) following the methodology in \citet{Max-Moerbeck2014} and \citet{lindfors2016optical}.

This result although not entirely consistent with the previous estimate is still within the same order of magnitude, also taking into account the fact that the delay found by these tests is not particularly prominent. 
From Fig. \ref{Fig:longterm}, we can roughly see that the light curves in the HE gamma rays and the optical-UV range also seem to follow these variability patterns, together with the optical R-band data, and similarly, the radio core (the surface where the jet becomes optically thin at this frequency) component seen in the VLBA (MOJAVE) data also follows the same variability pattern with the 15 GHz radio (OVRO) data. Because the seasonal gaps in the optical band are long ($\sim 150$ days), it is possible that such gaps could affect the result of the correlation test. We tested this by repeating the same analysis procedure for the radio data where the gaps are shorter (only $\sim 60$ days) and the HE gamma-ray data that has a weekly cadence. The power-law slope for generating the red-noise light curves for the HE gamma-ray data was 1.1. Correlating these data sets we found a time lag of $-140$ days $2\,\sigma$ confidence level (HE gamma-rays leading radio) and a standard deviation of 13 days. Similarly, we tested the correlation between the optical and the HE gamma-ray data and found only a minor lag of 20 days with $3\,\sigma$ confidence level (optical leading the HE gamma-rays) and a standard deviation of 12 days, confirming the physical origin of the radio-optical time lag. The existence of a such a time lag could mean that these emissions have a common origin, a blob or a shock travelling down the jet that is first emitting in the UV and optical bands in regions closer to the central engine. Further downstream where the jet becomes transparent to longer wavelengths, the particles originating from the emitting region would have lost their energy therefore emitting in the radio wavelengths.

Visually comparing the long-term variability in Fig. \ref{Fig:longterm} (notice the times of the two flares marked by the dashed lines) with the broadband changes seen in the light curve during the 2020 flare, it is possible to see that the enhanced emission of VER~J0521+211 appears only in HE and VHE gamma rays and in X-rays, but no increase in the emitted flux is observed in the radio and optical/UV bands during the VHE gamma-ray flare (see Fig. \ref{Fig:MWL_LC}). This suggests that the nature of the VHE gamma-ray flare in 2020 was substantially different from the one in 2013 presented in \citet{2Comp}, during which an enhanced flux was seen across all bands, from radio to VHE gamma rays. Assuming the validity of the two-component model that was described in \cite{2Comp}, the fact that the 2020 flare was detected only in high energies would imply a weaker contribution from the 'core' component (the source of lower energy emission) with respect to its contribution during the 2013 flare.

\subsection{Spectral analysis}
\label{sec:spectral analysis}

From the analysis of the nightly HE and VHE gamma-ray fluxes in Fig. \ref{Fig:MWL_LC}, it is possible to see that the emission in both bands features a dual-peaked pattern, with a first maximum between 25 and 26 February 2020 (MJD 58904-58905) and a second one in 1 March 2020 (MJD 58909), separated by a comparatively lower activity state. As shown in the next section, it is possible to define up to four states 
with significant differences in the emitted flux in both bands.

\subsubsection{VHE gamma rays}
\label{sec:res&vhe}

\begin{table*}
    \tiny
    \centering
    \caption{Results of the night-wise VHE gamma-ray spectral analysis for VER~J0521+211.}
    \tabcolsep=0.11cm
    \begin{tabular}{lccccccccccccc}
    \hline
    \hline
    State & Epoch  & T$_{\rm eff}$ &  Signif. & Model & F-test & Differential flux & $\Gamma$ & $\beta$ \\
     &   &  &   &  & & at 300 GeV ($\times 10^{-11}$) &  &  \\
       & [MJD]  &[h]& [$\sigma$] &  & & [$\rm cm^{-2} \rm s^{-1} \rm TeV^{-1} $] & & \\
    \hline   
       \textcolor{black}{\cite{2Comp}} & 56580-56627 &  4.50 & 30.47 &  LP & - & $27.43 \pm 0.51$ & $2.69 \pm 0.02 $ & $-0.47 \pm 0.07 $ \\
        \textcolor{green}{A} & 58903.5-58905.5 &  2.43 & 38.45 &  LP & $5.08\times 10^{-5}\ [\sim 4.8\sigma]$ & $5.12 \pm 0.31$ & $3.07 \pm 0.08 $ & $-1.47 \pm 0.28 $ \\
       \textcolor{blue}{B} & 58905.5-58906.5 &  0.66 & 10.34 & PWL & $1.38\times 10^{-1}\ [\sim 1.2\sigma]$ & $0.02 \pm 0.01 $ & $3.07 \pm 0.08 $ & - \\
       \textcolor{blue}{B}  & 58906.5-58907.5 &  0.65 & 11.58 & PWL & $6.95\times 10^{-2}\ [\sim 1.6\sigma]$ & $0.04 \pm 0.01 $ & $3.05 \pm 0.18 $ & - \\
        \textcolor{yellow}{C}   & 58907.5-58908.5 &  0.73 & 22.94 & LP & $1.25\times 10^{-5}\ [\sim 5.3\sigma]$ & $6.23 \pm 0.64$ & $3.35 \pm 0.23 $ & $-1.72 \pm 0.54 $ \\
       \textcolor{red}{D}   & 58908.5-58909.5 &  0.72 & 30.46 & LP & $1.18\times 10^{-3}\ [\sim 3.0\sigma]$ & $11.10 \pm 0.90$ & $3.04 \pm 0.12 $ & $-0.88 \pm 0.34 $ \\
       -   & 58909.5-58910.5 &  0.70 & 20.82 & PWL & $3.75\times 10^{-3}\ [\sim 2.5\sigma]$ & $0.22 \pm 0.03 $ & $2.69 \pm 0.10 $ & - \\
    \hline
    \end{tabular}
    \tablefoot{Columns: (1) State; (2) Observation epoch; (3) Effective time; (4) Detection significance; (5) statistically preferred model; (6) F-test probability (and equivalent significance expressed in $\sigma$); (7) Differential flux at 300 GeV; (8) and (9) Spectral index and curvature.
    Best-fit parameters are reported for the observed spectrum, without taking into account EBL absorption due to the uncertain redshift.}
    \label{tab:VHEspectra}
\end{table*}

As detailed in Sect. \ref{sec:magic}, MAGIC detected VER~J0521+211 on six consecutive nights. We report the individual nightly exposures and their corresponding detection significance in Table \ref{tab:VHEspectra}, and show the night-wise evolution of the VHE gamma-ray flux above 200 GeV in the top panel of Fig. \ref{Fig:MWL_LC}. The significant detection during each exposure enabled us to reconstruct the night-wise VHE gamma-ray spectra (Table \ref{tab:VHEspectra}). Additionally, we tested the spectral variability using both a power-law model (PWL; Eq. \ref{eq:PWL}) and a log-parabola model (LP; Eq. \ref{eq:LP}).

\begin{equation}
\frac{dF}{dE}(E)=F_0\left(\frac{E}{E_0}\right)^{-\Gamma}
\label{eq:PWL}
\end{equation}

\begin{equation}
\frac{dF}{dE}(E)=F_0\left(\frac{E}{E_0}\right)^{-\Gamma-\beta (\mathrm{log}_{10}(E/E_0))}
\label{eq:LP}
\end{equation}

\noindent 
where:


\begin{align*}
\tfrac{dF}{dE} =& \ \mathrm{differential \ flux} \\
F_0 =& \ \mathrm{normalisation \ constant} \\   
E_0 =& \ \mathrm{normalisation \ energy} \\ 
\Gamma =& \ \mathrm{photon \ index} \\ 
\beta =& \ \mathrm{curvature}\\
\end{align*}

We chose the model that describes the data best (see Table \ref{tab:VHEspectra}) using an F-test, defined as:

\begin{equation}
F = \frac{(RSS_1 - RSS_2)/(ndf_1 - ndf_2)}{RSS_2/ndf_2}
\end{equation}

In this expression, $RSS_{1,2} = \sum_{i=1}^n (y_i - f(x_i))^2$ is the residual sum of squares for each model and $ndf_{1,2}$ the corresponding number of degrees of freedom. We used a confidence level threshold of $3\,\sigma$, corresponding to a probability of about  $<2.7\times 10^{-3}$, to select the more complex (LP) over the simpler (PWL) model.

Table \ref{tab:VHEspectra} summarises the outcome of the spectral analysis for the VHE gamma-ray band. We identified four different states during the flaring episode based on the flux and spectral behaviour in the VHE gamma-ray band. The observations carried out on 27 and 28 February 2020 (MJD 58906 and  58907) have compatible spectral parameters and favour the same model with similar probabilities, therefore they were grouped as one state (B). For 02 March 2020 (MJD 58911), no MWL coverage is available to constrain the emission models, therefore we excluded this data from further analysis and interpretation.

\subsubsection{HE gamma rays}
\label{sec:res&he}

\begin{table}
    \tiny
    \centering
    \caption{Summary of the fit parameters to the {\it Fermi}-LAT spectra for each given epoch.}
    \tabcolsep=0.11cm
    \begin{tabular}{lccccccccccc}
    \hline
    \hline
      State & Epoch & TS  & Model &  $\Gamma _{\rm PWL}$  & F$_{0.1-300 \rm ~ GeV}$ \\
       & [MJD] & &  &   &     $[\times 10^{-7}\rm cm^{-2} \rm s^{-1} ]$\\
    \hline 
       \textcolor{green}{A} & 58903.5-58905.5 & 77 & PWL & $1.64 \pm 0.18$ &    $2.60 \pm 1.01 $ \\
       \textcolor{blue}{B} & 58905.5-58906.5 & 29  & PWL & $2.00 \pm 0.29$ &    $2.27 \pm 1.39 $ \\
       \textcolor{yellow}{C} & 58907.5-58908.5 & 55 & PWL & $1.47 \pm 0.28$ &    $2.46 \pm 1.27 $ \\
       \textcolor{red}{D} & 58908.5-58909.5 & 90 & PWL & $1.78 \pm 0.29$ &    $3.86 \pm 1.82 $ \\
    \hline
    \end{tabular}
    \tablefoot{ Columns: (1) State. (2) Observation epoch. (3) Test Statistics (TS). (4) Spectral model. (5) Spectral index. (6) Integral flux between $100\,\mathrm{MeV}$ and $300\,\mathrm{GeV}$.}
    \label{tab:HEspectra}
\end{table}

\begin{table*}
    \tiny
    \centering
    \caption{Summary of the fit parameters to the {\it Swift}-XRT spectra in each given epoch.}
    \begin{tabular}{lcccccccccccccccc}
    \hline
    \hline
      State & Time & T$_{\rm exp}$ &  OBS ID & Model & F-test & $\chi_{\rm red}^2$/ndf & $\Gamma _{\rm LP}$  & F$_{2-10\rm ~keV}$ & F$_{0.3-10\rm ~keV}$ \\
      \cline{9-10}
      & [MJD]  &[ks]& &    & [\%] & & &  \multicolumn{2}{c}{[$\times 10^{-11} \rm erg ~ cm^{-2} \rm s^{-1} $]}\\
    \hline    
     \textcolor{green}{A} & 58904.12262  & 1.04 & 00031531058 &  LP & 0.02 & $0.97/266$ & $2.23 \pm 0.09$  
      & $4.30 \pm 0.34$ & $6.25 \pm 0.38$ \\
      \textcolor{blue}{B} & 58906.36604  & 1.43 & 00031531061 & LP & 0.03 & $1.16/88$ & $2.15 \pm 0.05$  
      & $4.57 \pm 0.20$ & $6.43 \pm 0.22$ \\
      \textcolor{yellow}{C} & 58907.84000  & 0.59 & 00031531064 & LP & 0.14 & $0.87/196$ & $2.32 \pm 0.11$  
      & $4.84 \pm 0.53$ & $7.29 \pm 0.50$ \\
      \textcolor{red}{D} & 58908.83296  & 0.90 & 00031531063 & LP & 0.00 & $0.79/242$ & $2.26 \pm 0.10$  
      & $5.84 \pm 0.54$ & $8.57 \pm 0.55$ \\
    \hline
    \end{tabular}
    \tablefoot{Columns: (1) State. (2) Time of the observation. (3) Exposure. (4) Swift observation ID. (5) Preferred spectral model. (6) F-test (LP over PWL). (7) Reduced $\chi^2$ and number of degrees of freedom for the LP. (8) Spectral index. (9) Integral flux between $2\,\mathrm{keV}$ and $10\,\mathrm{keV}$. (10) Integral flux between $0.3\,\mathrm{keV}$ and $10\,\mathrm{keV}$.}
    \label{tab:XRAYspectra}
\end{table*}

Simultaneous monitoring of \textit{Fermi}-LAT during the acquisition of MAGIC data resulted in significant detection of the source for each night except for the two nights of state B, which have lower fluxes compared to the other states (see Sect. \ref{sec:res&vhe} and Fig. \ref{Fig:MWL_LC}). The HE gamma-ray flux is similar for 25 and 26 February 2020 (MJD 58903-MJD 58905). Therefore, we defined state A to cover both nights. Table \ref{tab:HEspectra} summarises the best-fit model parameters (for the PWL case, Eq. \ref{eq:PWL}) for each of the four states and the corresponding test statistics (TS).

\subsubsection{X-rays}
\label{sec:res&xr}

\textit{Swift}-XRT observations were performed contemporaneously to MAGIC data for the four defined states (see Fig. \ref{Fig:MWL_LC}) within a window of about $12\,\mathrm{h}$, and almost simultaneous with MAGIC for states C and D. VER~J0521+211 is a bright X-ray source, but due to its location near the Galactic plane, a strong hydrogen absorption is expected to occur at energies below $\sim 1 \mathrm{\,keV}$. The intrinsic spectrum, corrected for such absorption, is well described by a log-parabolic model (Eq. \ref{eq:LP}). We report the best-fit parameters in Table \ref{tab:XRAYspectra}. When comparing the different analysis states we see no significant changes in the spectral shape, but the X-ray flux during state D was roughly $20\%$ higher compared to previous states.

\subsection{Redshift estimation}
\label{sec:res&zest}

\begin{figure}
\centering
\includegraphics[width=1\linewidth]{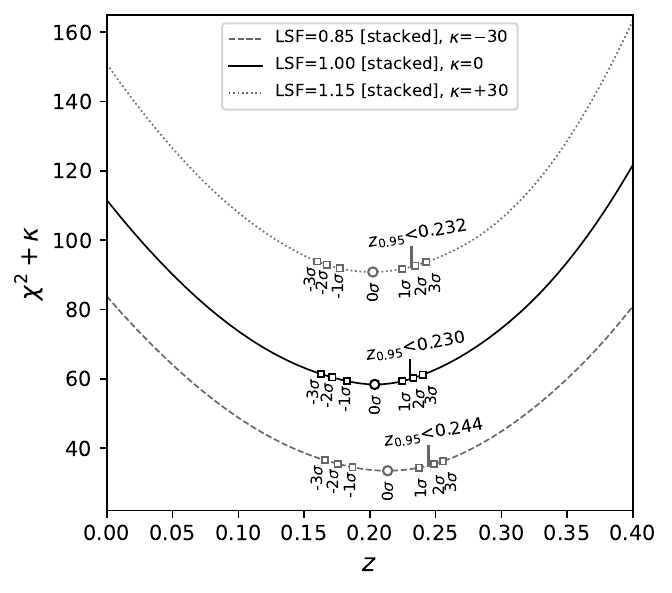}
\caption{Statistical redshift reconstruction using a profile-$\chi^2$ scan for different light scaling factors, jointly for the four states considered in this work. The solid, dotted, and dashed curves represent the profile-$\chi^2$ for a given light scaling factor (nominal, and $\pm 15\%$ respectively). The best-fit value is shown for each case as an open circle, the $\pm 1\,\sigma$, $\pm 2\,\sigma$, and $\pm 3\,\sigma$ two-sided confidence bands represented as small open squares. Finally, the $95\%$ confidence level upper limit is represented as a vertical bar and its actual value shown in the plot. An artificial constant $\kappa$ was added to two of the curves for clarity purposes.}
\label{Fig:z_UL}
\end{figure}

The determination of a precise distance to VER~J0521+211 remains, as mentioned in Sect. \ref{sec:intro}, elusive. A proposed lower limit for its redshift, provided by \citet{paiano2017redshift} at $z \gtrsim 0.18$, is quoted in several recent works. This estimation relies on an assumption on the (non-detected) host galaxy, $M_R = -22.9$. \citet{paiano2017redshift} also discuss the possibility of having a dimmer blazar, with absolute magnitude $M_R = -21.9$. The lack of detection of the host galaxy would modify the lower limit to just $z \gtrsim 0.10$. The complexity to measure the redshift of VER~J0521+211 arises from its featureless optical spectrum. Nonetheless, important advances have been made in providing strong upper limits to the redshift based on certain assumptions on the shape of the intrinsic spectrum of the source in gamma rays. \citet{veritas2022multiwavelength} determined a statistical upper limit to the redshift of $z \lesssim 0.31$ assuming that the intrinsic VHE gamma-ray spectrum is not harder than the HE gamma-ray spectrum.

In this work, we followed the methodology described in \citet{acciari2019_multiebl} to derive upper limits for the redshift using a likelihood ratio test and taking into account instrumental uncertainties. The method assumes a concave log-parabola as the spectral model to describe both the HE and VHE gamma-ray data simultaneously, and we treated each of the four states (A--D) independently to produce a profile likelihood (in this case a profile-$\chi^2$) as a function of redshift. In a second step, we combined the profile likelihood for the four states (see Fig. \ref{Fig:z_UL}). For this procedure, we considered both the constraints to the shape of the spectrum from VHE gamma-ray data (fitting the number of source and background events for each energy bin in our VHE gamma-ray data with the spectral model folded by the instrument response function) and from {\it Fermi}-LAT data (using exclusively the values and uncertainties on the flux and spectral index at the pivot energy, where both parameters become least correlated). The selection of different extragalactic background light (EBL) models reported in \cite{acciari2019_multiebl} revealed that the uncertainties arising from the choice of EBL model are subdominant with respect to instrumental calibration and reconstruction uncertainties, mainly regarding the energy scale. Hence, we used only the model described in \citet{dominguez2011extragalactic} to determine a 95\% confidence level upper limit on the source redshift. We estimated the instrumental calibration uncertainties by scaling the simulation of the total light collected by the instrument, applying a light scaling factor (LSF) in $\pm$15\% increments, and performing independent reconstruction with the corresponding modified instrument response functions. Out of this analysis, we obtained the worst-case limits when the Cherenkov light is scaled up in the simulations by $15\%$, yielding a 95\% confidence level upper limit of $z \leq 0.244$ for the combined profile-$\chi^2$, as shown in Fig. \ref{Fig:z_UL}. This limit is among the best upper limits to date, and it is consistent with the lower limit of $z \gtrsim 0.18$ reported in \citet{paiano2017redshift}.  Consequently, we assumed a redshift value of $z$ = 0.18 in all the following analyses, particularly during the modelling of the broadband SED of the source (see also results in Sect. ~\ref{sec:sed&modelling}).

\section{Spectral energy distribution modelling}
\label{sec:sed&modelling}

Broadband SED modelling is a powerful tool to investigate the origin of the emission in blazars. One-zone emission models are the simplest among the models we can use. They have comparatively few free parameters, and are especially descriptive when the emissions across different wavelengths correlate well, suggesting a common origin \citep[see e.g.][for previous studies of VER~J0521+211]{archambault2013discovery,2Comp,veritas2022multiwavelength}. However, during the 2020 flaring episode, short-term variability was predominantly observed at high energies, including X-rays and beyond, suggesting a complex emission region structure not reproducible with a simple one-zone model. Moreover, while one-zone models can adequately describe the optical-to-VHE-gamma-ray part of the broadband SED, they often overlook the radio band \citep{Tavecchio2016} which is correlated (with a time lag) with the optical emission in the case of VER J0521+211 (see Sect. \ref{sec:res&mwlvar}), indicating a common origin for these energies. Therefore, in this study we opted for a two-zone leptonic model instead. Previously, \cite{2Comp} used a two-component model to describe the SED of VER~J0521+211. In that work, they assumed the two-zones to be co-spatial and interacting, mimicking a spine-sheath model. However, as discussed in Sect. \ref{sec:res&mwlvar}, the time lag between optical and radio emissions suggests that lower-energy radiation is emitted later, farther down the jet, thus supporting the non-interaction between the two components.

While leptonic jet models usually suffice to describe the SED of many BL Lacs, the possibility of blazars as sources of neutrinos \citep[e.g.,][]{Petropoulou2017,Petropoulou2020,Petropoulou2020b,IceCube2018,Kreter2020} has prompted interest in models with a hadronic component. In the case of a purely leptonic scenario, one-zone models have often been ruled out based on the rationale mentioned earlier. However, lepto-hadronic one-zone scenarios have proven effective in modelling some SEDs. This scenario assumes that the lower-energy part of the spectrum results from synchrotron emission by electrons and positrons, while the high-energy part arises from proton synchrotron and photo-hadronic interactions, in addition to the leptonic inverse Compton that is present in the purely-leptonic case \citep[see e.g.,][]{Cerruti2011}. 

\begin{figure*}
\centering
\includegraphics[width=0.8\linewidth]{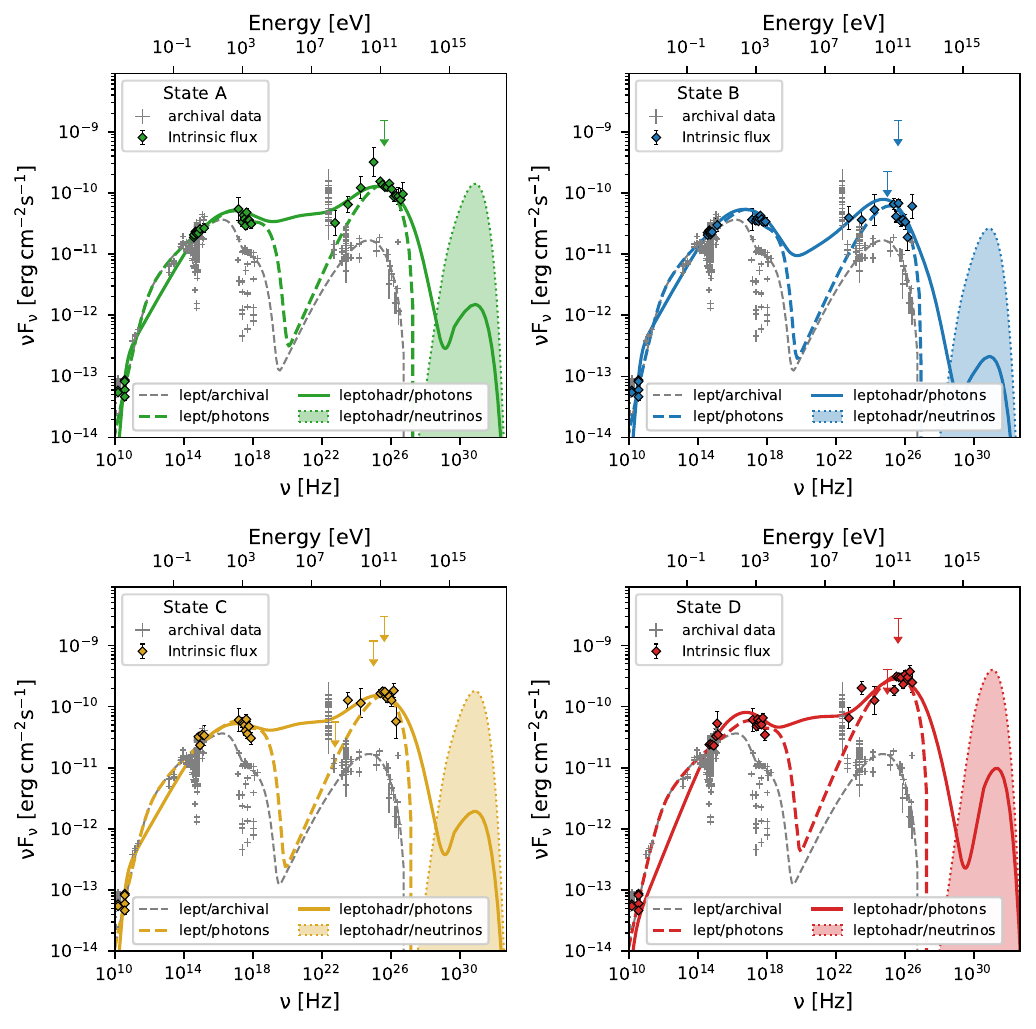}
\caption{Intrinsic broadband SEDs of the four observation epochs listed in Tab.~\ref{tab:VHEspectra}. Archival data are shown in grey \citep{2Comp}, while the data contemporary to the four epochs are shown in their respective  colours. The fluxes predicted by the two-zone leptonic model are shown dashed, while the predictions of the one-zone lepto-hadronic model are shown as solid lines.In both cases, the models are shown to match the intrinsic flux measured by MAGIC, without taking into account $\gamma$-$\gamma$ absorption by extragalactic background light (relevant for TeV photons) and cosmic microwave background (completely absorbing PeV photons). The spectra of emitted neutrinos predicted by the lepto-hadronic model are shown as dash-dotted lines.}
\label{Fig:SED_TEvo_Core}
\end{figure*}

Figure~\ref{Fig:SED_TEvo_Core} illustrates the time-resolved broadband SEDs, together with the best-fitting two-zone leptonic and one-zone lepto-hadronic models, which we describe more in detail in the following subsections. In all cases, we fixed the redshift of VER~J0521+211 to the lower limit suggested in \cite{paiano2017redshift}, that is $z\geq 0.18$.

\subsection{Two-zone leptonic model (non-interacting)}
\label{sec:SEDmodelling_2znon}

The light curve of the 2020 flare (Fig.~\ref{Fig:MWL_LC}) exhibits two distinct trends, as described in Table~\ref{tab:VHEspectra} and discussed in Sect.~\ref{sec:res&mwlvar}. On one hand, the gamma-ray data from {\it Fermi}-LAT and MAGIC display a double-peaked pattern with an intermediate state in between. In contrast, the X-ray emission shows a steady but elevated state, which later during the maximum of the VHE emission culminated on a single peak in the emitted flux. Furthermore, the UV, optical, and radio fluxes remained at a quiescent level, and no significant changes in the polarisation degree or strong rotations of the EVPA were detected around the time of the flare (see Fig. \ref{Fig:optpolfit}). Given that no significant changes in the radio band have been seen after the expected 160-day time lag either, the physical interpretation arguably needs to be different than that of the previous flares.

To model this flare, we constructed a framework consisting of two separate non-interacting spherical regions: a 'blob' (smaller, higher energy region closer to the black hole) and a 'core' (larger, lower energy region further out in the jet). Both regions are filled with relativistic electrons characterised by a broken power-law energy distribution. In contrast to the model in \cite{2Comp} where the two regions were co-spatial and interacting, the time lag between the radio and optical bands suggests that some separation between the jet emitting regions exists especially when the higher energy emission should originate closer to the central engine. Therefore, we assumed that no significant interaction or photon feedback between the two populations of energetic particles exists. Additionally, we considered the 'self-absorption' effects resulting from pair production in the region further out negligible.

Building upon the best-fit parameter set found for the core component (spectral indices and minimum and break Lorentz factors) in \cite{2Comp} and taking into consideration the fundamental differences between the two models (i.e. two-zone interacting in \cite{2Comp}, two-zone non-interacting in this work), we developed the four two-zone models, trying to manually find solutions that modify the values of as few parameters as possible between two consecutive states. These four solutions effectively capture the time-evolution of the broadband SED from radio frequencies to the VHE gamma-ray band, as depicted in Fig.~\ref{Fig:SED_TEvo_Core}.

The corresponding model parameters are shown in Table~\ref{tab:SEDmodpar}. Notably, such a simple two-zone model successfully reproduces the evolving spectra for the four phases, requiring only variations in the parameters of the electron spectrum, magnetic field intensity B, and electron density $N*$. 

The breakdown of the two-zone leptonic model SED in individual emission processes is available in Appendix \ref{sec:app_2zoneleptonic}. 

\begin{table*}
    \tiny
    \centering
    \caption{Parameters of the non-interacting two-zone model.}
    \tabcolsep=0.11cm
    \begin{tabular}{lccccccccccccccccc}
    \hline
    \hline
      State & Epoch & Model & $R$ & $\delta$ & $B$ & $n_1$ & $n_2$ & $\gamma_\mathrm{min}$ & $\gamma_\mathrm{b}$ & $\gamma_\mathrm{max}$ & $N^*$ & $\mathrm{L_e}$ & $\sigma$ & $z$ & $\chi^2$\\
       & [MJD]  & (region) & [$\times10^{15}\ \mathrm{cm}$] &  & [$\mathrm{G}$] & & & [$\times10^3$] & [$\times10^4$] & [$\times10^6$] & [$\times 10^3 \ \mathrm{cm}^{-3}$] & [$\mathrm{erg/s}$]& & & (points)\\
    \hline
    \textcolor{green}{A}   & 58903.5-58905.5 &  2-zone (blob) & 13 & 11 & 0.10 & 1.70 & 3.4 & 1.00 & 25 & 2.0 & 1.2 & $1.8\times 10^{44}$ &  $4.3\times 10^{-3}$  & 0.18 & 152.2\\
      &  &  2-zone (core) & 370 & 11 & 0.13 & 1.64 & 2.7 & 0.35 & 0.11 & 0.2 & 2.5 & $6.2\times 10^{43}$ & 16 & 0.18 & (33)\\
    \hline  
    \textcolor{blue}{B}   & 58905.5-58907.5 &  2-zone (blob) & 13 & 11 & $\boldsymbol{0.20}$ & 1.70 & 3.4 & 1.00 & $\boldsymbol{10}$ & $\boldsymbol{1.0}$ & $\boldsymbol{0.8}$ & $\boldsymbol{8.2\times 10^{43}}$  & $\boldsymbol{3.7\times 10^{-2}}$& 0.18 & 334.9\\
      &  &  2-zone (core) & 370 & 11 & 0.13 & 1.64 & 2.7 & 0.35 & 0.11 & 0.2 & 2.5 & $6.2\times 10^{43}$ & 16 & 0.18 & (30)\\
    \hline  
    \textcolor{yellow}{C}   & 58907.5-58908.5 &  2-zone (blob) & 13 & 11 & $\boldsymbol{0.11}$ & 1.70 & 3.4 & 1.00 & $\boldsymbol{30}$ & $\boldsymbol{1.5}$ & $\boldsymbol{0.11}$ & $\boldsymbol{1.7\times 10^{44}}$ & $\boldsymbol{5.4\times 10^{-3}}$& 0.18 & 24.6 \\
      &  &  2-zone (core) & 370 & 11 & 0.13 & 1.64 & 2.7 & 0.35 & 0.11 & 0.2 & 2.5 & $6.2\times 10^{43}$  & 16& 0.18 & (23)\\
    \hline
    \textcolor{red}{D}   & 58908.5-58909.5 &  2-zone (blob) & 13 & 11 & $\boldsymbol{0.08}$ & 1.70 & 3.4 & 1.00 & $\boldsymbol{30}$ & 2.0 & $\boldsymbol{2.0}$ & $\boldsymbol{3.1\times 10^{44}}$  & $\boldsymbol{1.6\times 10^{-3}}$& 0.18 & 266.1\\
      &  &  2-zone (core) & 370 & 11 & 0.13 & 1.64 & 2.7 & 0.35 & 0.11 & 0.2 & 2.5 & $6.2\times 10^{43}$  & 16& 0.18 & (28)\\
    \hline
    \end{tabular}
    \tablefoot{Columns: (1) State. (2) Epoch. (3) Model (emission region). (4) Radius of the emission region. (5) Doppler factor. (6) Magnetic field strength. (7) and (8) Slopes of the electron spectrum around $\gamma_\mathrm{b}$. (9) Minimum electron Lorentz factor. (10) Break electron Lorentz factor. (11) Maximum electron Lorentz factor. (12) Electron density. (13) Electron luminosity. (14) Jet magnetisation given as the ratio of the Poynting luminosity \citep{Celotti_2008MNRAS.385..283C} to the total particle luminosity $L_\mathrm{B}/L_\mathrm{e}$. (15) Redshift. (16) Sum of squared residuals normalized by the uncertainties on the flux points ($\chi^2$), excluding the radio band. The number of flux points is reported in parenthesis. In bold are parameters that changed compared to state A.}
    \label{tab:SEDmodpar}
\end{table*}

\subsection{One-zone lepto-hadronic model}
\label{sec:SEDmodelling_1z}

\begin{table*}
    \tiny
    \centering
    \caption{Parameters of the lepto-hadronic one-zone model.}
    \begin{tabular}{lcccccccccccccccc}
    \hline
    \hline
      State & Epoch & R$_\mathrm{blob}$ & $\delta$ & $B$ & $n_\mathrm{e}$& $\gamma_\mathrm{e}^\mathrm{min}$  & $\gamma_\mathrm{p}^\mathrm{min}  $  &$L_\mathrm{e}$ [erg/s] & $\sigma$ & $\chi^2$\\
     & [MJD] &   [$\times 10^{16}$ cm]& & [G]  & $n_\mathrm{p}$ & $\gamma_\mathrm{e}^\mathrm{max}$ &   $\gamma_\mathrm{p}^\mathrm{max}$ & $L_\mathrm{p}$ [erg/s] & & (points) \\
    \hline
    \textcolor{green}{A} &
   58903.5-58905.5 &  2.3 & 15.0 & 0.15 & 2.0 & $1.3 \times 10^1$ &  $6.31 \times 10^5$ & $1.8\times 10^{42}$ & $8.3\times 10^{-5}$ & 548.4 \\
    & &   &  &  & 2.0 & $6.3 \times 10^5$ &  $5.0 \times 10^7$ &  $3.0\times 10^{46}$ & & (33) \\

    \hline  
     \textcolor{blue}{B}  & 58905.5-58907.5 &  2.3 & 15.0 & 0.15 & 2.0 & $1.3\times 10^1$ &  $6.31 \times 10^5$ & $\boldsymbol{2.2\times 10^{42}}$  & $\boldsymbol{5.0 \times 10^{-4}}$ & 248.7\\
    & &   &  &  & 2.0 & $6.3 \times 10^5$ &  $5.0 \times 10^7$ &  $\boldsymbol{5.0\times 10^{45}}$ & & (30) \\

    \hline  
     \textcolor{yellow}{C}  & 58907.5-58908.5 &  2.3 & 15.0 & 0.15 & 2.0 & $1.3\times 10^1$ &  $6.31 \times 10^5$ & $\boldsymbol{1.8\times 10^{42}}$ & $\boldsymbol{6.6\times 10^{-5}}$ & 26.3\\
    & &   &  &  & 2.0 & $6.3 \times 10^5$ &  $5.0 \times 10^7$ &  $\boldsymbol{3.8\times 10^{46}}$ & & (23)\\

    \hline
      \textcolor{red}{D} & 58908.5-58909.5 &   2.3 & 15.0 & 0.15 & \textbf{1.6} & $\boldsymbol{2.5\times 10^1}$ &  $\boldsymbol{5.0\times 10^6}$ & $\boldsymbol{1.3 \times 10^{42}}$& $\boldsymbol{1.4\times 10^{-4}}$  & 238.0\\
    & &   &  &  & \textbf{1.7} & $\boldsymbol{2.0\times 10^5}$ & $\boldsymbol{6.3\times 10^7}$ &  $\boldsymbol{1.8\times 10^{46}}$ & & (28) \\
    \hline
    \end{tabular}
    \tablefoot{Columns: (1) State. (2) Epoch. (3) Blob radius. (4) Doppler factor of the blob. (5) Magnetic field strength. (6) Power-law spectral indices for electron and proton distribution. (7) Minimum and maximum energy for electrons. (8) Minimum and maximum energy for protons. (9) Electron and proton luminosity. (10) Jet magnetisation given as the ratio of the Poynting luminosity \citep{Celotti_2008MNRAS.385..283C} to the total particle luminosity $L_\mathrm{B}/(L_\mathrm{p}+L_\mathrm{e})$. (11) Sum of squared residuals normalized by the uncertainties on the flux points ($\chi^2$), excluding the radio band. The number of flux points is reported in parenthesis. In bold are parameters that changed compared to state A.}
    \label{tab:lephad}
\end{table*}

In this section, we demonstrate that the four observed activity states of the source during the flare can also be accounted for within the one-zone lepto-hadronic framework. To explore this scenario, we employed the time-dependent code AM$^3$~\citep{gao2017direct}, which simulates interactions among accelerated protons, electrons, and the jet. AM$^3$ uses numerical techniques to solve the system of differential equations governing the evolution of the particle and photon spectra, ensuring a fully time-dependent and self-consistent treatment. We consider the one-zone leptohadronic model as an alternative to the two-zone leptonic with the same hypothesis of mostly leptonic origin of gamma rays. \cite{2015MNRAS.450L..21Z} and \cite{2020ApJ...893L..20L} showed that for low-frequency and intermediate-frequency BL Lacs the hadronic contribution to the gamma ray emission can only be subdominant. Since during the periods of high activity VER~J0521+211 is a HBL, a purely hadronic origin of high energy emission during these states is not excluded. Thus, a model where the gamma rays originate from proton synchrotron emission is possible. However, we do not consider proton synchrotron model in this paper and leave it for future work.

In our model, we assumed that electrons and protons follow simple power-law spectra, $dN/d\gamma_\mathrm{e(p)} \propto\gamma_\mathrm{e(p)}^{-n_\mathrm{e(p)}}$, with spectral index $n_\mathrm{e(p)}$, spanning a range of Lorentz factors from  $\gamma_\mathrm{e(p)}^\mathrm{min}$ to $\gamma_\mathrm{e(p)}^\mathrm{max}$. Both particle populations are then isotropically injected into a single spherical region of size $R$ (in the comoving frame of the jet), where a homogeneous and isotropic magnetic field of strength $B$ exists. Along with electron and proton luminosities, $L_\mathrm{e}$ and $L_\mathrm{p}$, this defines a ten-parameter space that fully describes the model. We chose to fix the Doppler factor of the blob at 15 (and 11 in the leptonic case) based on the assumption made on the viewing angles in \cite{2Comp}. For each emission state, we considered the entire multiwavelength dataset. 

We started the modelling from the state A, assuming it originates from a steady-state emission. After finding the best-fit solution with a genetic algorithm, we further refined it locally using the Minuit package \citep{1975CoPhC..10..343J}. Given the short time difference between consecutive days, of one or two days, we did not treat them as entirely independent models. Instead, we aimed to find the minimal parameter changes between two consecutive states that can account for the observed spectral changes. We fixed in Minuit all parameters to a constant value except one at a time, and iteratively minimise the reduced $\chi^2$ function for each of the parameters. If a satisfactory fit was not achieved, we explored combinations of two or more parameters to be freed during the minimisation. 

Our analysis revealed that the transition from state A (steady-state) to state B (enhanced) can be explained by an increase of electron luminosity by $20\%$, and a decrease of proton luminosity by a factor of six. Similarly, the transition from state B to C required only adjustments in electron and proton luminosities. In particular, the ratio between proton to electron luminosity increased again during state C, reaching a value similar to that during state A. However, this smooth transition is broken when reaching state D, which cannot be explained by variations in only two parameters. A satisfactory fit of the model to state D was only obtained by modifying each particle energy spectra along with their luminosities. 

In our model, both particle species are assumed to be already pre-accelerated at the moment of injection into the emission zone. Most particle acceleration mechanisms predict the same relative energy gain for both electrons and protons if they were accelerated co-spatially. We note that for the states A, B, and C the energy spectra of the particles do not change. The changes of the particle luminosities in these states can be interpreted as variations of the amount of particles that enter the emission region which can be caused, for example, by the turbulences of the magnetic field in the jet. State D, however, requires a significant change in both particle spectra and luminosities. Such a change can be caused by the combination of cooling of the particle populations from the previous state and simultaneous injection of new more energetic particles.

The best-fit SED models for each of the four states are represented in Fig. \ref{Fig:SED_TEvo_Core}, showing a good overall agreement with the data. The parameters of this one-zone model in which protons and electrons are co-accelerated, are shown in Table \ref{tab:lephad}. The presented model attributed the bulk of the low-energy emission (radio to optical) to synchrotron radiation from accelerated electrons, whereas the emission detected by MAGIC stemmed mainly from hadronic interactions between accelerated protons and low-energy photon fields in the jet. The emission detected by {\em Fermi}-LAT at GeV energies represents a transitional region, where the emission shifts from electron-dominated to proton-dominated.

The minimum electron Lorentz factors for all states are similar, $\sim 10$. Such low-energy electrons are responsible for the radio emission in our model. They do not contribute to the high energy part of the SED as the inverse Compton effect can upscatter their synchrotron photons only up to the optical and UV band which is dominated by the synchrotron emission of high-energy electrons (see Fig. \ref{Fig:SED_comp}). The contribution of these low-energy electrons to the hadronic processes is negligible as well. They can be treated as a separate electron population, thus, being compatible with the electrons from the core in the two-zone model. We note that if the low-energy electrons are located in the same zone as the high-energy electrons, the radio data should only be considered as an upper limit, implying that the minimal Lorentz factor in our best fit should be interpreted as a lower limit.

The model predicted well the emission of high-energy neutrinos through interactions between photons and protons, primarily produced during the decay of secondary charged pions (see Appendix \ref{sec:app_leptohadronic} for more details on the subject). The predicted neutrino flux is represented by the dash-dotted lines in Fig. \ref{Fig:SED_TEvo_Core}. Since neutrinos follow the energy distribution of the accelerated protons, their expected flux in the observer's frame has a maximum around 400 PeV. To estimate the number of neutrinos that the IceCube observatory would detect from the source during each of the four states, we integrated the emitted all-flavour neutrino flux spectra over the entire duration of states A, B, C, and D. We convolved the resulting fluence with the IceCube effective area for the source's declination band \citep{Aartsen_2017}, and divided it by three to account for neutrino mixing during propagation and detection, which only uses one channel. The calculations yield predictions of $0.010$, $0.002$, $0.013$, and $0.016$ neutrino events per day during the respective states, summing up $\sim 0.05$ neutrino events during the whole 6-day observation window of MAGIC. Considering Poisson statistics, these numbers are consistent with the absence of neutrino detections reported by \citet{IceCube:2019cia}. Therefore, the current upper limits on the neutrino flux from the IceCube experiment do not rule out the lepto-hadronic scenario. However, as the IceCube observatory and future neutrino telescopes continue to monitor the neutrino sky, they might eventually provide a detection which would support the lepto-hadronic emission scenario, particularly if the source is detected again during a flaring state in the VHE gamma-ray regime.

Similarly to the purely leptonic scenario, we present the breakdown of the lepto-hadronic model in Appendix \ref{sec:app_leptohadronic}. One difference with the leptonic case is the value adopted for the Doppler factor $\delta$, that is 15 for the lepto-hadronic case as opposed to 11 for the leptonic case. It should be noted however that $\delta$ is not well constrained for VER~J0521+211, as the jet opening angle is not measured \citep[See details in:][]{2Comp}. A second notable difference with the leptonic case is how the model fits the energy range between the two major peaks (synchrotron and inverse-Compton). In the one-zone lepto-hadronic model the dip between the two emission peaks almost completely disappears whereas in the two-zone leptonic model we see the typical two-bump structure and a much deeper dip in between. Observations from UV to X-rays, for example with NuSTAR, would be especially important in order to characterise the MWL spectrum in more detail. In addition, the proposed lepto-hadronic model only attempts to explain the varying emission from VER~J0521+211 during the 6-day high state observed by MAGIC. The introduction of a steady core component in the leptonic scenario stemmed from the fact that a single-zone leptonic scenario cannot convincingly reproduce the observed broadband SED \citep[see e.g.][]{2Comp}. No such study has been performed so far with a lepto-hadronic scenario. Furthermore, the inclusion of additional components on the latter, while potentially improving constraints on possible non-varying components present in the jet, would likely result in a significantly more complex (and degenerate) model to constrain, beyond the scope of this work.

\section{Optical polarisation modelling}
\label{sec:optpol}

As pointed out in Sect. \ref{sec:res&mwlvar}, we see a clear difference in the behaviour of the 2013 and the 2020 flares (see Fig. \ref{Fig:longterm}) where the former is seen in all wavelengths whereas the latter is only seen in wavelengths from X-rays to VHE gamma rays. This lead us to conclude that likely the long-term conditions of the jet have changed between these flaring episodes. 
In order to asses the possible change of jet conditions in the lower energies, we modelled the long-term optical polarisation signatures of VER 0521+211 to see if any change could be found.

Optical polarisation signatures observed in blazars have historically been attributed to two components. The typical scenario is a constant core component and a variable jet component \citep{valtaoja1991,villforth2010,barresdealmeida2010,barresdealmeida2014}. The need for the two components in polarisation modelling stems from the often drastic changes observed in the polarisation signatures, namely the large amplitude changes in the polarisation degree (PD) and the rotations observed in the electric vector polarisation angle (EVPA). These changes cannot be accounted for by a uniform, ordered magnetic field; instead, disturbances like shocks or turbulence are needed to create disorder in the magnetic field of the jet plasma.

For this study, we built on previous work presented in \citet{2Comp} where the polarisation signatures of multiple blazars were modelled using optical R-band data collected with the NOT telescope between November 2014 and November 2018 (MJD 56970-58430). The goal of that work was to determine the flux ratio of stable (or slowly varying) optical emission and the observed varying optical emission by simultaneously modelling optical intensity, PD and EVPA. The authors estimated the prior range for the stable optical component by analysing both optical and radio light curves following the methodology in \citet{lindfors2016optical}. The variable component was modelled as a cylindric emission region contained within a jet, with a helical magnetic field, and the changes seen in the modelled polarisation were expected to arise from the change between the flux ratio of the constant and the variable emission components. The calculation of the Stokes parameters was based on the formulae described in \citet{luytikov2005}. Their model successfully constrained the range of the stable emission component for some sources, but in the case of VER J0521+211 the posterior errors entirely filled the prior range of this parameter.

In this work, we introduced a more physical model for the variable component. Instead of assuming that the total intensity tracks directly the density inside the emission cylinder, we explicitly modelled the density variations as described below and compute the Stokes parameters the total intensity $I$, $Q$, and $U$. We added more data (as detailed in Sect. \ref{sec:pol}) to guarantee a better coverage of the polarisation signatures over time.

In our model, the emission in the optical band originates from the stable core component that is analogous to the 'core' of the SED modelling although with some modifications to its geometry. The density of the core is disturbed by individual, moving structures (called 'spheres' from here on). The optical polarisation variability is then measured only within the volume of the core. The core component corresponds to a cylindrical region with dimensions: radius $R$, length $l$, and hot plasma flowing through it at velocity $\beta c$. The plasma carries a helical magnetic field with strength $B$ and pitch angle $\phi$ ($\phi=90^{\circ}$ meaning magnetic lines perpendicular to the jet axis and $\phi=0^{\circ}$ corresponding to the parallel case). The relativistic flow of the plasma forms a viewing angle $\theta$ with the line of sight and an angle $\eta$ with the projection of the jet on the plane of the sky. Thus in this case, the core component resembles a thin disk instead of a more spherical emission region. The core component, filled with a homogeneous electron population of density $K$, is described by eight parameters in total. In order to see whether the current jet conditions could explain the long-term polarisation variability, we used the results from the SED modelling (see Sect. \ref{sec:sed&modelling} and Table \ref{tab:polresults_core}) to fix the following input parameters: Doppler factor $\delta$ (and thus the jet velocity $\beta$), magnetic field strength $B$, and emission region radius $R$. $l$ is calculated from the half-life of the electrons emitting in R-band in the observer frame.

\begin{figure*}
\centering
\includegraphics[width=1.0\textwidth]{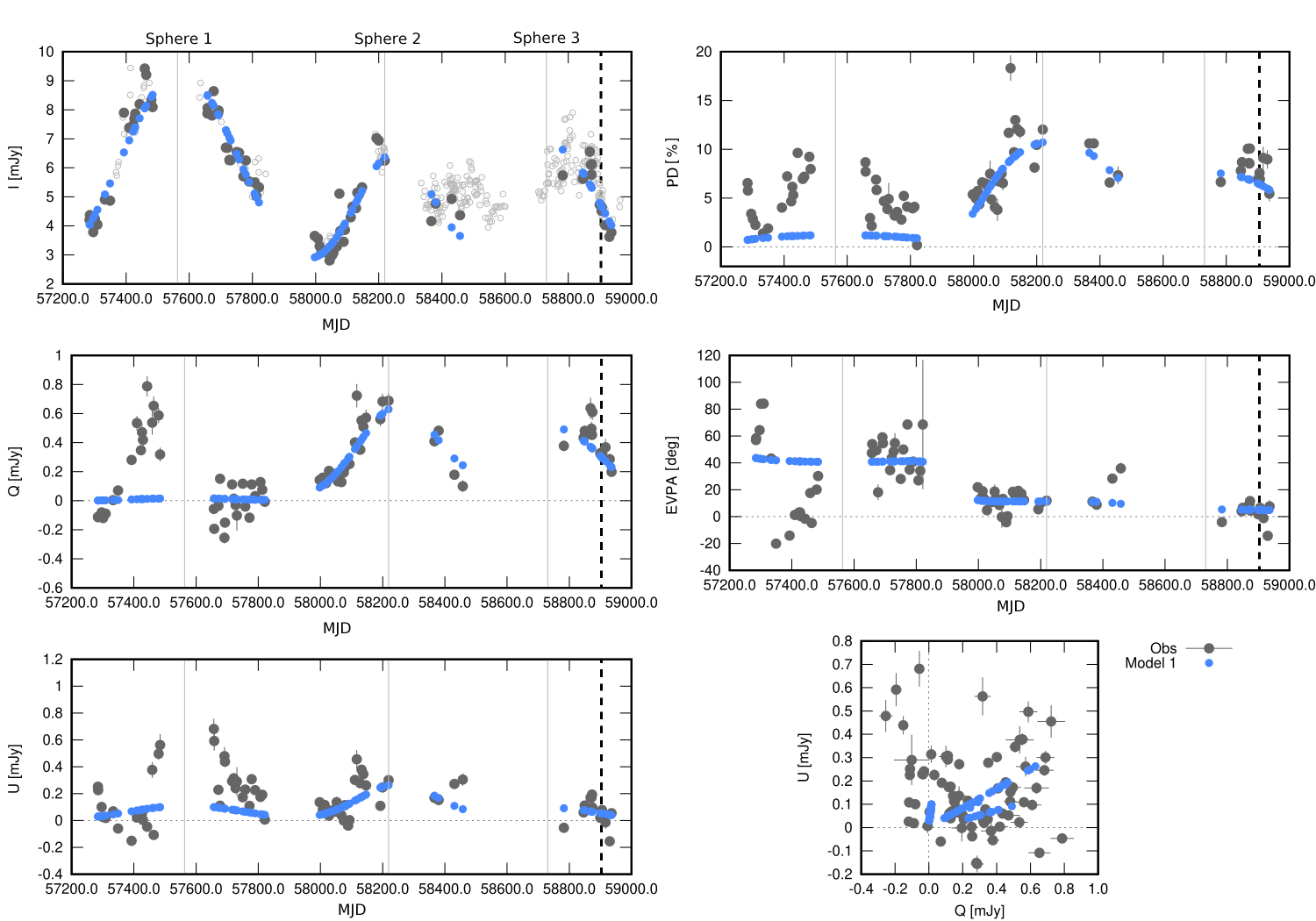}
\caption{
Modelling results (blue circles) of the optical (R-band) polarisation data (grey circles). The small open circles denote optical intensity data that has no simultaneous polarisation observations. Such data are excluded from the modelling. Panels from top to bottom on the left: optical intensity, Q and U Stokes parameters. Panels from top to bottom on the right: polarisation degree, electric vector position angle, Stokes parameters Q-U plane. Grey vertical lines indicate epochs corresponding to the entrance time of the spherical density enhancements (spheres) into the emission zone. Dashed vertical lines indicate the time of the 2020 high-energy flare.}
\label{Fig:optpolfit}
\end{figure*}

The variable component is described by spherical density enhancements with a Gaussian profile. These spheres cross the cylindrical core emission region causing the observed outbursts, notably around MJD 57530, 58230, and 58810 (see Fig. \ref{Fig:optpolfit} where each outburst is marked with vertical lines during the time of the sphere arrival times). Each sphere is described by five parameters: entry time $T_0$, characteristic radius $R_{\rm sphere}$ defined at $1\,\sigma$ of the Gaussian profile, the distance from the sphere to the centre of the cylinder $D_{\rm sphere}$, the pitch angle $\phi_{\rm sphere}$ with respect to the cylinder  axis, and $A_{\rm sphere}$ the density ratio between each sphere and the core (see Table \ref{tab:polresults_spheres}). The model calculates the variability only based on the part of the sphere embedded in the cylinder. We fit the model to the data using a downhill simplex walker method, incorporating an evolutionary approach over $\sim 40-50$ generations to find the global minimum due to the method's susceptibility to local minima. Only epochs with both photometry and polarimetry data were included in the fit, ensuring a balanced influence.

The results of the polarisation modelling are summarised in Fig. \ref{Fig:optpolfit}, which shows the predicted values of $I$, PD, EVPA, Stokes parameters $Q$ and $U$, and the Q-U plane as a function of time, compared against the real measurements. Tables \ref{tab:polresults_core} and \ref{tab:polresults_spheres} display the corresponding best-fit parameters for the core and the three sphere components of the model. Compared to the model describing the broadband SED, the electron density of the core component differs by a factor of six. This is probably due to a combination of different emission region sizes (spherical instead of a thin disk) and slightly different electron spectrum assumptions between the two models. Notably, the model predicts $D_{\rm sphere}/R > 1$ for the three major outbursts caused by each sphere, suggesting that the centre of the sphere moves {\em outside} the core emission cylinder. Since we consider only the density enhancements inside the cylinder, the variations in the model result from their movement around the edges of the cylinder. In this scenario, these enhancements can thus break the symmetry of the density profile and ultimately cause an increase in the polarisation degree.

Out of the three outbursts observed in Fig. \ref{Fig:optpolfit}, the latter two outburst (Spheres 2 and 3) are well-described by the model as seen in each panel. The first outburst (Sphere 1) is characterised by fast EVPA rotations, up to $100^\circ$, which are not reproduced by our model and therefore only the intensity variations are adequately reproduced. We could assume that the underlying physics in the jet was different during the first outburst epoch after which the jet conditions might have changed before the latter two outbursts, explaining why the model is unable to reproduce all three outbursts simultaneously. In fact, Fig. \ref{Fig:longterm} illustrates a clear increase in the intensity of the VLBI core component tracing the long-term optical and radio trends, implying global changes in the jet that affect the optical polarisation. These changes could also explain the different nature of the 2013 and the 2020 VHE gamma-ray flares (as seen in Fig. \ref{Fig:longterm}).

Comparing these results to those of the SED modelling, we found a good consistency between the two-zone leptonic model and the polarisation model, but notably found a discrepancy in the electron densities of the core component of each model. In order to alleviate this, the volume of the cylinder in the polarisation model could be increased by a factor of 21 that is the ratio of the emission region volumes of the two models. Then the density of the core would be $15/21 \sim 0.7 \ \mathrm{cm^{-3}}$. Taking into account the fact that only the part of the sphere inside the cylinder contributes to the model (estimating it to be 5\% of the peak enhancement) we can obtain an electron density $ K = 0.7\times 85\times 0.05 \sim 3 \ \mathrm{cm^{-3}}$ that is very close to the value in the two-zone SED model ($n_e \sim 2$). However, increasing the emission region volume to obtain this would, in turn, compromise one of the previously fixed values, either $R$ (fixed from the SED modelling) or $l$ (calculated from the half-life of the electrons emitting in R-band in the observer frame). We also point out that whereas in the SED modelling both the core and the blob are seen contributing to the optical emission, in the polarisation modelling we only considered optical emission originating from the core region. In Sect. \ref{sec:res&mwlvar}, we found a time lag of 160 days between the radio and optical light curves, which would suggest some separation between the emission regions of these two bands. Therefore, it seems likely that to fully reproduce the optical polarisation signatures a more complex geometry is needed.

Although this model represents a substantial advancement in understanding the jet geometry via optical polarisation, we acknowledge that it does not test a polarisation model with a single component. Therefore, while the model fits the data adequately, it does not exclude other possible interpretations. Finally, to cause rotations in the Q-U plane that would be observed as rotations of the EVPA, the spheres likely need to rotate along a helical magnetic field or there needs to be a phase effect of several components.

\begin{table}
    \tiny
    \centering
    \caption{Results of the polarisation modelling for the core component.}
    \begin{tabular}{lccccccccccc}
    \hline
    \hline
      $\beta$ & $\theta$ & $B$ & $\phi$ & $\mathrm{log}(R/\mathrm{cm})$ & $\mathrm{log}(l/\mathrm{cm})$ & $K$ & $\eta$  \\
       & [$^{\circ}$] & [$\mathrm{G}$] & [$^{\circ}$] &  &  & [cm$^{-3}$] & [$^{\circ}$]  \\
    \hline
    0.98 & 0.4 & 0.13 & 29 & 17.56 & 16.38 & 15 & 206 \\
    \hline
    \end{tabular}
    \tablefoot{Columns: (1) Jet velocity as a fraction of $c$. (2) Jet viewing angle. (3) Magnetic field strength (fixed from SED fitting). (4) Magnetic field pitch angle. (5) Radius of the emission region (fixed from SED fitting). (5) Length of the emission region cylinder. (6) Electron density. (7) Projected angle of the jet in the sky.}
    \label{tab:polresults_core}
\end{table}

\begin{table}
    \tiny
    \centering
    \caption{Results of the polarisation modelling for each of the three spherical components (column 1).}
    \begin{tabular}{lccccccccc}
    \hline
    \hline
      Component & $T_0$ & $R_{\rm{sphere}}$ & $D_{\rm{sphere}}$ & $\phi _{\rm{sphere}}$ & $A_{\rm{sphere}}$ \\
      & [MJD] &  &  & [$^{\circ}$] &   \\
    \hline
    Sphere 1 & 57559.7 &  1.3 & 1.7 & $-68$  & 33 \\
    Sphere 2 & 58278.6 &  0.9 & 2.1 & $-183$ & 92 \\
    Sphere 3 & 58769.0 &  1.1 & 2.3 & $187$  & 80 \\
    \hline
    \end{tabular}
    \tablefoot{ Columns: (2) Time of sphere's entry into the emission region. (3) Sphere radius in units of emission cylinder radius $R$ (see Table \ref{tab:polresults_core}). (4) Distance of the sphere measured from the centre of the emission region, same unit as for $R_{\rm sphere}$. (5) Position angle of the entry point. (6) Relative density of the sphere, expressed in multiples of $K$ (see Table \ref{tab:polresults_core}).}
    \label{tab:polresults_spheres}
\end{table}

\section{Summary and conclusions}
\label{sec:sum&con}

We presented a study of the short-term evolution of the emission from the BL Lac object VER~J0521+211 during a flaring state, which occurred between February and March 2020. A collection of instruments observed the blazar, providing comprehensive coverage across the electromagnetic spectrum. This MWL campaign spanned six consecutive nights that we divided into four distinct states. To put the observations in context, we gathered long-term broadband measurements of the source from 2009 to 2021, as well as optical R-band polarimetry data from 2014 to 2021.

We examined the variability of the source across all wavelengths during the 2020 campaign for the flare. Notably, we observed no variability in the optical and radio frequencies, with flux levels consistent with previous archival data. This contrasts with a previous detection at VHE gamma rays of the source reported in \cite{2Comp}, where the flare was observed across all wavelengths, even though in the radio band the flaring state was seen several months later. This difference hints at a potential change in the internal structure or configuration of the jet, modifying the physical mechanisms responsible for the flare. To further support this, we examined the polarimetry data, where we found no significant changes in polarisation degree or EVPA during the period around the flare. Focusing on the long-term picture, our study revealed a time lag of $-160\pm 11.1$ days between radio and optical bands, with the optical leading the flux changes. This delay is of a similar magnitude with previous estimations \citep{lindfors2016optical,2Comp}, although better constrained due to the use of more data. We interpret this as a potential common origin for the lower energy emissions although these emissions are not entirely co-spatial.

The redshift of VER~J0521+211 is still unknown as of today, due to the lack of features in its optical spectrum. Recent observations have focused on providing statistical lower and upper limits for the redshift estimation, based respectively on the non-detection of its host galaxy and the extrapolation of the HE emission to the VHE band, or modelling of gamma-ray data. Using the new data from MAGIC presented in this work, along with contemporaneous data from \textit{Fermi}-LAT in the HE band, we were able to improve the existing statistical upper limits on the redshift to $z \leq 0.244$ at $95\%$ confidence level. The new estimation of the redshift of VER~J0521+211, incorporating the host-luminosity-dependent lower limit from \citep{paiano2017redshift}, is therefore set to $0.18 \leq z \leq 0.244$.

We explored two emission scenarios to model the flare of 2020: a purely leptonic model with two non-interacting emission regions, and a lepto-hadronic one-zone scenario. The modelling was performed for four distinct states during the flare, varying smoothly only a few parameters at a time between consecutive states. In both frameworks, the proposed models successfully capture the time-evolving broadband SEDs of VER~J0521+211 across a wide frequency range, spanning from radio to VHE gamma rays. By incorporating hadronic interactions in the lepto-hadronic scenario, the model extended the predicted photon emission to frequencies of $\sim 10^{31}\,\mathrm{Hz}$, where the strong gamma-ray opacity from photon-photon interactions with the cosmic microwave background (CMB) will make it impossible to detect it. This VHE gamma-ray radiation, attributed to neutral pion decay, is a byproduct of the neutrino emission from VER~J0521+211 predicted by our lepto-hadronic model, which current and future neutrino instruments may be able to probe. Furthermore, the two models describe the emission between the two major emission components differently (the UV-X-ray dip) and more detailed observations especially in this energy range are necessary to characterise the MWL SED shape and differentiate between these two models. Alternatively, the detection of significant X-ray polarisation with IXPE, or future missions like COSI, would set unprecedented constraints on the possible hadronic component in VER~J0521+211.

Finally, our analysis of the long-term MWL behaviour of the source revealed that the flaring activity behaviour of the sources changes through time. In the archival data, we see that flaring is seen almost simultaneously in all wavelengths, whereas in the case of the 2020 flare, higher activity occurs only in the higher energies, X-rays and beyond. In order to understand the reason behind this phenomenon, we modelled the long-term optical polarisation data. The model was based on the work presented in \cite{2Comp}, but introduced a more physical scenario. In this picture, outbursts are produced when spherical density enhancements enter into the constant core component. In order to make the results more comparable, we set fixed values for the Doppler factor, magnetic field strength, and the emission region radius based on the results of the leptonic SED modelling. The long-term optical polarisation light curve featured three significant flares, with our model effectively reproducing the latter two but not the first one. Such a discrepancy likely supports the idea of a change of the physical processes affecting the jet over time. Therefore, while successful, the results of our polarisation modelling leave room for more complex models that could eventually explain the entire evolution of the polarisation of the emission from the source.

\begin{acknowledgements}

M. Artero: MAGIC and Fermi-LAT data analysis, redshift estimation, publication coordination; V. Fallah Ramazani: Principal investigator, MAGIC, optical, Swift-XRT, and radio data analysis, coordination of multiwavelength observations, publication coordination, interpretation, modelling discussion, J. Jormanainen: MAGIC, optical, and radio data analysis, coordination of multiwavelength observations, publication coordination, interpretation, polarisation modelling, redshift estimation; E. Lindfors: interpretation and discussion of polarisation and leptonic SED modelling; M. Nievas Rosillo: Swift-UVOT and Swift-XRT data analysis, publication coordination, interpretation, leptonic SED modelling, redshift estimation; K. Nilsson: interpretation, polarisation modelling; A. Omeliukh: lepto-hadronic SED modelling; X. Rodrigues: lepto-hadronic SED modelling. The rest of the authors have contributed in one or several of the following ways: design, construction, maintenance, and operation of the instrument(s); preparation and/or evaluation of the observation proposals; data acquisition, processing, calibration and/or reduction; production of analysis tools and/or related Monte Carlo simulations; discussion and approval of the contents of the draft.

We would like to thank the Instituto de Astrof\'{\i}sica de Canarias for the excellent working conditions at the Observatorio del Roque de los Muchachos in La Palma. The financial support of the German BMBF, MPG and HGF; the Italian INFN and INAF; the Swiss National Fund SNF; the grants PID2019-104114RB-C31, PID2019-104114RB-C32, PID2019-104114RB-C33, PID2019-105510GB-C31, PID2019-107847RB-C41, PID2019-107847RB-C42, PID2019-107847RB-C44, PID2019-107988GB-C22, PID2022-136828NB-C41, PID2022-137810NB-C22, PID2022-138172NB-C41, PID2022-138172NB-C42, PID2022-138172NB-C43, PID2022-139117NB-C41, PID2022-139117NB-C42, PID2022-139117NB-C43, PID2022-139117NB-C44 funded by the Spanish MCIN/AEI/ 10.13039/501100011033 and “ERDF A way of making Europe”; the Indian Department of Atomic Energy; the Japanese ICRR, the University of Tokyo, JSPS, and MEXT; the Bulgarian Ministry of Education and Science, National RI Roadmap Project DO1-400/18.12.2020 and the Academy of Finland grant nr. 320045 is gratefully acknowledged. This work was also been supported by Centros de Excelencia ``Severo Ochoa'' y Unidades ``Mar\'{\i}a de Maeztu'' program of the Spanish MCIN/AEI/ 10.13039/501100011033 (CEX2019-000920-S, CEX2019-000918-M, CEX2021-001131-S) and by the CERCA institution and grants 2021SGR00426 and 2021SGR00773 of the Generalitat de Catalunya; by the Croatian Science Foundation (HrZZ) Project IP-2022-10-4595 and the University of Rijeka Project uniri-prirod-18-48; by the Deutsche Forschungsgemeinschaft (SFB1491) and by the Lamarr-Institute for Machine Learning and Artificial Intelligence; by the Polish Ministry Of Education and Science grant No. 2021/WK/08; and by the Brazilian MCTIC, CNPq and FAPERJ. This research has made use of data from the OVRO 40-m monitoring program (Richards, J. L. et al. 2011, ApJS, 194, 29), supported by private funding from the California Institute of Technology and the Max Planck Institute for Radio Astronomy, and by NASA grants NNX08AW31G, NNX11A043G, and NNX14AQ89G and NSF grants AST-0808050 and AST- 1109911. This publication makes use of data obtained at the Mets\"ahovi Radio Observatory, operated by the Aalto University. The research at Boston University was supported in part by the National Science Foundation grant AST-2108622,  and by several NASA Fermi Guest Investigator grants, the latest is 80NSSC23K1507. This study was based in part on observations conducted using the 1.8m Perkins Telescope Observatory (PTO) in Arizona, which is owned and operated by Boston University. This research was partially supported by the Bulgarian National Science Fund of the Ministry of Education and Science under grants KP-06-H38/4 (2019) and KP-06-PN-68/1 (2022). This research has made use of data from the MOJAVE database that is maintained by the MOJAVE team (Lister et al. 2018). V.F.R. was supported by the Academy of Finland projects 317636, 320045, 346071, and 322535. J.J. was supported by the Academy of Finland projects 320085, 322535, and 345899, as well as by the Alfred Kordelin Foundation. E.L. was supported by the Academy of Finland project Nos. 317636, 320045 and 346071. M.N.R. acknowledges the funding support from the Severo Ochoa program, the Agencia Española de Investigación (Ministerio de Ciencia, Innovación y Universidades), and the European Union. A.O. was supported by DAAD funding program 57552340. X.R. was supported by the Deutsche Forschungsgemeinschaft (DFG, German Research Foundation) under Germany's Excellence Strategy – EXC 2094 – 390783311.
\end{acknowledgements}

%
%
\bibliographystyle{aa} 
\bibliography{Draft_VER_J0521+211} 

\begin{appendix}
\renewcommand\thefigure{\thesection.\arabic{figure}}

\section{Breakdown of the leptonic SED}
\label{sec:app_2zoneleptonic}

\begin{figure}
\centering
\includegraphics[width=.47\textwidth]{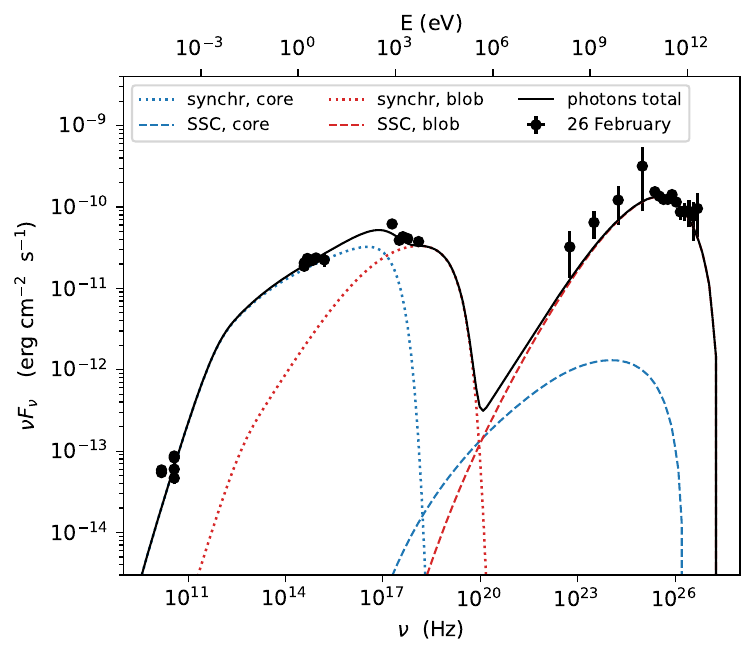}
\caption{Contributions from different radiative processes to the SED of VER~J0521+211 (state A) in a leptonic scenario, where two non-interacting regions (blob and core) are responsible for the total emission detected. For the core, the synchrotron and SSC emission are represented by dotted and dashed blue lines, while for the blob, these radiation fields are shown in dotted and dashed red curves respectively. The total emission is shown as a black curve.}
\label{fig:SEDlep_comp}
\end{figure}

A detailed leptonic model for state A is shown in Fig. \ref{fig:SEDlep_comp}. This model involves two non-interacting leptonic components: the blob and the core. When we look at the energy spectrum, these components are clearly distinguishable. The core's synchrotron component is responsible for most of the light emitted from radio to soft X-rays (below 1 keV), while the blob's synchrotron radiation takes over beyond 1 keV. In the proposed model, the majority of gamma-ray emission comes from the blob's synchrotron self-Compton (SSC) mechanism, surpassing the core's SSC emission by up to two orders of magnitude.

\section{Breakdown of the lepto-hadronic SED}
\label{sec:app_leptohadronic}

\begin{figure}
\centering
\includegraphics[width=.46\textwidth]{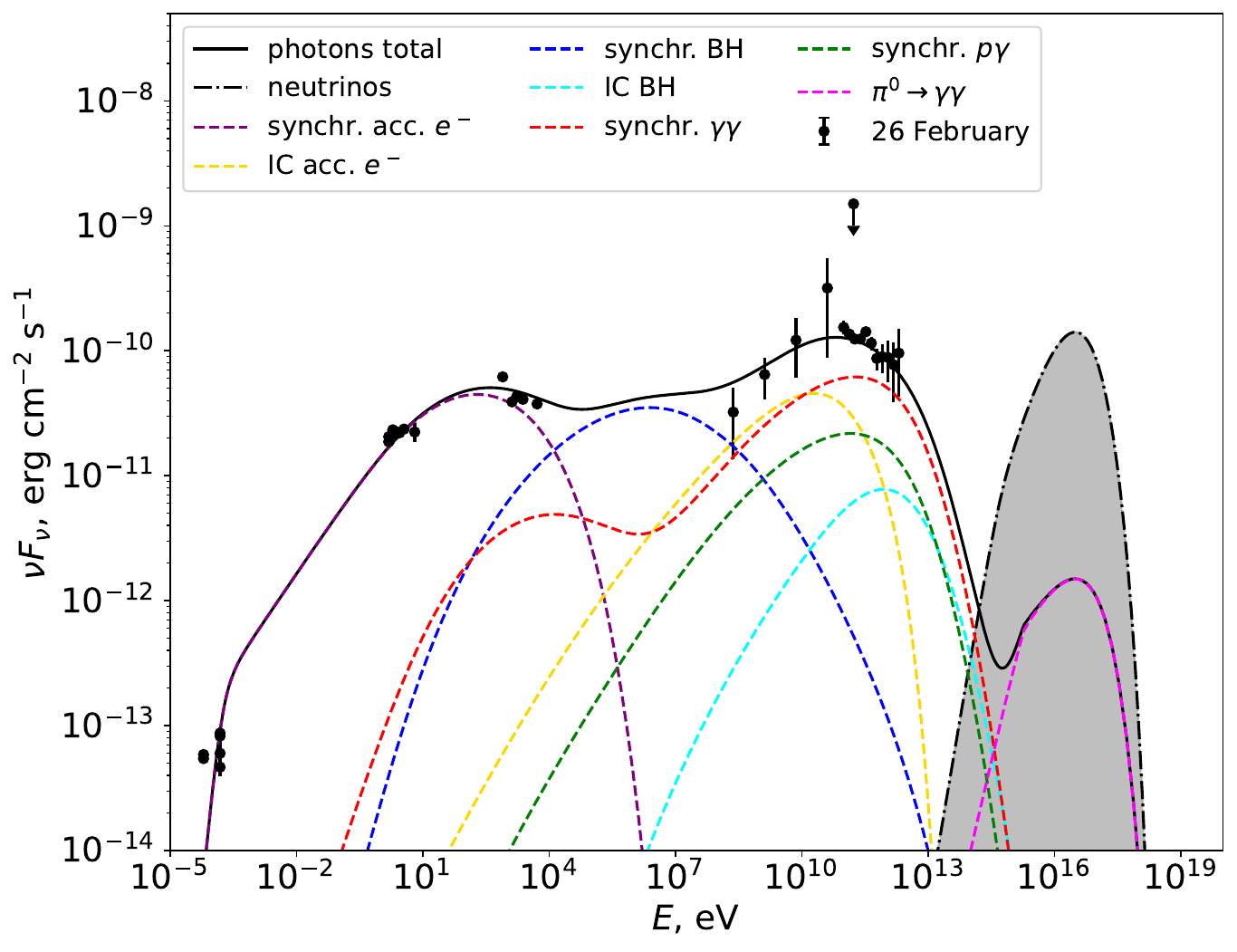}
\caption{Contributions from different radiative processes to the SED of VER~J0521+211 (state A), in a lepto-hadronic scenario where protons are co-accelerated to $\gtrapprox10~\mathrm{PeV}$. The solid line shows a total radiated photon flux, same as in Fig. \ref{Fig:SED_TEvo_Core} (upper left panel). Dashed lines correspond to the following contributions: purple -- synchrotron radiation from electrons accelerated in the jet, blue -- synchrotron radiation from pairs created in Bethe-Heitler process, orange -- SSC emission, red -- synchrotron radiation from $\gamma \gamma$ pair creation, green -- synchrotron from electrons created in hadronic cascades following p$\gamma$ interactions, cyan -- inverse Compton Bethe-Heitler electrons, and magenta -- pion decay to $\gamma\gamma$. Neutrinos are shown with dash-dotted filled curve.}
\label{Fig:SED_comp}
\end{figure}

From the comparison with the two-zone leptonic model, we notice that the shape of SED in the one-zone lepto-hadronic model is different from that in the two-zone leptonic model. Instead of the characteristic two-bump feature, the dip in X-rays almost disappears when adding protons. Figure \ref{Fig:SED_comp} shows the contributions from leptonic and hadronic processes to the total photon fluxes.

The low-energy emission is explained almost exclusively by the synchrotron emission of the electrons accelerated in the jet. Those electrons also comptomise their synchrotron radiation, resulting in high-energy contribution (orange dashed line in Fig. \ref{Fig:SED_comp}). Another leptonic process that contributes significantly to the GeV -- TeV gamma rays is synchrotron emission from pair production $\gamma \gamma \leftrightarrow e^{+} e^{-}$.

The emission from co-accelerated protons becomes dominant over leptonic emission in the energy range 10 keV -- 0.1 GeV. The main contribution in this energy range (blue dashed line in Fig. \ref{Fig:SED_comp}) comes from synchrotron emission of electrons created in Bethe-Heitler process $p\gamma \rightarrow p e^{+} e^{-}$. At the higher energies (10 GeV -- 100 TeV), synchrotron emission of electrons from hadronic cascades in $p \gamma$ interactions ($p \gamma \rightarrow \pi^{\pm} \rightarrow \mu^{\pm} \rightarrow e^{\pm}$) and inverse Compton scattering of synchrotron radiation of Bethe-Heitler electrons have subdominant effect. A separate bump around $10^{17}$ eV (magenta dashed line in Fig. \ref{Fig:SED_comp}) comes from the direct $\pi^0 \rightarrow \gamma \gamma$ decays and thus those photons are produced via a hadronic process. However, this ultra-high-energy component is severely attenuated by photon-photon interactions with CMB photons.
\end{appendix}

\end{document}